\documentclass[12pt]{article}

\newlength{\dinwidth}
\newlength{\dinmargin}
\usepackage{amsthm, amsfonts, bm, mathrsfs, indentfirst, sectsty,
graphicx, fancyhdr, slashed, fullpage, color, amssymb,caption,subcaption}

\usepackage{hepth}

\usepackage[colorlinks=true,linktocpage=true,linkcolor=blue,citecolor=blue]{hyperref}

\usepackage{authblk}

\theoremstyle{definition}

\theoremstyle{remark}

\def\s{\sigma}

\def\beq{\begin{equation}}
\def\enq{\end{equation}}

\newcommand{\be}{\begin{equation}} \newcommand{\ee}{\end{equation}}
\newcommand{\bea}{\begin{eqnarray}} \newcommand{\eea}{\end{eqnarray}}
\newcommand{\beann}{\begin{eqnarray*}}  \newcommand{\eeann}{\end{eqnarray*}}
\newcommand{\bfig}{\begin{figure}} \newcommand{\efig}{\end{figure}}
\newcommand{\ba}{\begin{array}} \newcommand{\ea}{\end{array}}
\newcommand{\bcen}{\begin{center}} \newcommand{\ecen}{\end{center}}
\newcommand{\btab}{\begin{tabular}} \newcommand{\etab}{\end{tabular}}

\newcommand{\matt}{\left ( \begin{array}{ccc}}
\newcommand{\ematt}{\end{array} \right )} \newcommand{\matf}{\left (\begin{array}{cccc}}
    \newcommand{\ematf}{\end{array} \right )} \newcommand{\vect}{\left (
\begin{array}{c}}
    \newcommand{\evect}{\end{array} \right )}    \def\beqn{\begin{eqnarray}}
 \def\eeqn{\end{eqnarray}}  
     
\newtheorem{Proposition}{Proposition}[section]

\newtheorem{Theorem}{Theorem}[section]
\newtheorem{Lemma}{Lemma}[section]
\newtheorem{Corrolary}{Corrolary}[section]

\newcommand{\bp}{\begin{Proposition}}   \newcommand{\ep}{\end{Proposition}}
\newcommand{\bt}{\begin{Theorem}}   \newcommand{\et}{\end{Theorem}}
\newcommand{\bl}{\begin{Lemma}}     \newcommand{\el}{\end{Lemma}}
\newcommand{\bc}{\begin{Corrolary}} \newcommand{\ec}{\end{Corrolary}}

\begin{document}

\begin{titlepage}
\hfill MCTP-14-16

\hfill MPP-2014-311


\begin{center}

{\Large \bf Holographic p-wave Superconductor with Disorder}

\vspace{.5cm}

{\bf \small D. Are\'an$^a$, A. Farahi$^b$,  L. A. Pando Zayas$^b$, I. Salazar Landea$^{c}$ and A. Scardicchio$^d$}

\vspace{.4cm}

{\it ${}^a$ Max-Planck-Institut f\"ur Physik (Werner-Heisenberg-Institut)}\\
{\it F\"ohringer Ring 6, D-80805, Munich, Germany }\\

\vspace{.3cm}

{\it ${}^b$ Michigan Center for Theoretical Physics}\\
{\it Randall Laboratory of Physics, University of Michigan}\\
{\it  Ann Arbor, MI 48109, USA}

\vspace{.3cm}

{\it ${}^c$ Instituto de F\'\i sica La Plata (IFLP) and Departamento de F\'\i sica}\\
{\it   Universidad Nacional de La Plata, CC 67}\\
{\it  1900 La Plata, Argentina}\\

\vspace{.3cm}
{\it ${}^{c,d}$ International Centre for Theoretical Physics (ICTP)}\\
{\it   Strada Costiera 11, I 34014 Trieste, Italy }\\

\vspace{.3cm}
{\it ${}^{d}$ Physics Department, Princeton University, Princeton, NJ 08542, USA}\\
\vspace{.3cm}
{\it ${}^{d}$ Physics Department, Columbia University, New York, NY 10027, USA}\\
\vspace{.3cm}
{\it ${}^{d}$ ITS, Graduate Center, City University of New York, New York, NY 10016, USA}\\
\vspace{.3cm}
{\it ${}^{d}$ INFN, Sezione di Trieste, Strada Costiera 11, 34151, Trieste, Italy}

\vspace{12pt}

\end{center}
\begin{abstract}

We implement the effects of disorder on a holographic p-wave
superconductor by introducing a random chemical potential  which defines the local energy of the charge carriers.
Since there are various possibilities for the orientation of the vector order parameter,
we explore the behavior of the condensate in the parallel and perpendicular directions to the introduced disorder.
We clarify the nature of various branches representing competing solutions and construct the disordered phase diagram.
We find that moderate disorder enhances superconductivity as determined by the value of the condensate. Though we mostly focus on uncorrelated noise, we also consider a disorder characterized by its spectral properties and study in detail its influence on the spectral
properties of the condensate and charge density. We find fairly universal responses of the resulting power spectra
characterized by linear functions of the disorder power spectrum.
\end{abstract}

\end{titlepage}

\tableofcontents

\setcounter{page}{1} \renewcommand{\thefootnote}{\arabic{footnote}}
\setcounter{footnote}{0}

\hrulefill


\section{  Introduction}

The  AdS/CFT correspondence \cite{Maldacena:1997re,Witten:1998qj,Gubser:1998bc,Aharony:1999ti} provides a window into the dynamics of strongly coupled systems by  identifying the underlying field theory with a weakly coupled gravity dual. In recent years the methods and scope of the AdS/CFT correspondence have shifted from traditionally QCD-motivated problems to problems in the area of condensed matter systems (see reviews \cite{Hartnoll:2009sz,Herzog:2009xv,McGreevy:2009xe,Sachdev:2010ch}, and references therein). In particular, various models of holographic s-wave \cite{Hartnoll:2008vx,Hartnoll:2008kx} and p-wave \cite{Gubser:2008zu,Gubser:2008wv} superconductors have been constructed.

Among the various paradigms in condensed matter physics, disorder is a fundamental one as it provides a crucial step away from clean systems toward realistic ones. One striking  manifestation of disorder in non-interacting quantum systems is the phenomenon of Anderson localization \cite{Anderson:1958vr}, where the conductivity can be completely suppressed by quantum effects. The study of the interplay between disorder and interactions in quantum systems has seen little progress on the theoretical side. Recently, however, in the context of disordered conductors, Basko, Aleiner and  Altshuler presented compelling evidence in favor of a many-body localized phase, based on an analysis of the perturbation theory in electron-electron interaction to all orders \cite{Basko20061126}. Subsequent works (see \cite{PhysRevB.75.155111,PhysRevB.82.174411,PhysRevB.84.094203,0295-5075-101-3-37003} and references therein) have confirmed and sharpened the existence
of a phase transition separating the weakly and strongly interacting limits of electrons in disordered potentials.

Disorder is also particularly relevant in the context of superconductors; it has a rich history dating back to the pioneering work of Anderson in 1959 \cite{Anderson195926}. For many years Anderson's theorem, stating that superconductivity is insensitive to perturbations that do not destroy time-reversal invariance (pair breaking), provided the central intuition. Critiques to Anderson's argument were raised, for example, in \cite{doi:10.1143/JPSJ.51.1380,PhysRevLett.54.473,PhysRevB.33.3146,PhysRevB.32.5658} where the effects of strong localization were considered. More generally, the interplay between interactions and disorder in superconductors cannot be considered settled. In view of this situation, it makes sense to consider alternative models where the problem can be analyzed in full detail.

Indeed, in a previous work \cite{Arean:2013mta}, we initiated a program of directly studying the role of disorder in
holographic superconductors which arguably apply to strongly interacting superconductors. There have been other approaches
to disorder in holography
\cite{Hartnoll:2007ih, Hartnoll:2008hs,Fujita:2008rs,Ryu:2011vq,Adams:2011rj,Adams:2012yi,Saremi:2012ji}.
As in \cite{Arean:2013mta}, we follow a very direct approach to the realization of disorder by coupling an operator to a randomly distributed space-dependent source. Essentially, we directly translate a typical condensed matter protocol into the AdS/CFT framework. Namely, we choose a random space-dependent chemical potential by setting the boundary value of a $U(1)$ electric potential. The main rationale for this choice of disorder relies on the fact that the chemical potential defines the local energy of a charge carrier placed at a given position $x$, as it couples to the particle number  $n(x)$ locally. Therefore, our choice of disorder
replicates a local disorder in the on-site energy.  This is the simplest protocol one would implement.
Moreover, once disorder is introduced in such an strongly interacting system, all observables will become disordered and,
therefore, the physics is not expected to depend on the way disorder is originally introduced.

The direct approach outlined above has now been applied by other authors in the context of holography.
 For example, it was used to argued for Anderson localization in \cite{Zeng:2013yoa}. Other interesting
 applications  include \cite{Hartnoll:2014cua} and  \cite{Lucas:2014zea}.

It is worth mentioning that another very important motivation for our work is related to the more general and far-reaching problem of translational invariance in holography. Most holographic models respect translational invariance in the field theory directions. This  underlying translational
invariance has adverse effects in applications involving transport properties in condensed matter. Since translational invariance implies momentum conservation, it  means that the charge carriers have nowhere to dissipate their momentum, resulting
in a zero frequency delta function in the optical conductivity which obscures interesting questions
such as the temperature dependence of the DC resistivity. A lot of effort has recently been devoted to addressing this
shortcoming. Some progress has been reported in  \cite{Hartnoll:2012rj,Sonner:2013aua,Herzog:2014tpa}.
Another approach  to momentum
 dissipation include models of massive gravity \cite{Vegh:2013sk}, \cite{Blake:2013owa},
 \cite{Amoretti:2014zha,Amoretti:2014mma}.
However, this latter approach struggles with issues of UV completeness of the gravity models used.

The paper is organized as follows. In section \ref{Sec:ReviewPW}, we review the construction of the holographic p-wave
superconductor. In section \ref{Sec:Disorder} we introduce our implementation of disorder and present some typical results.
In section \ref{Sec:Thermo} we describe the different branches that emerge in our setup and determine which one wins the
thermodynamic competition by comparing the free energy. In section \ref{Sec:Phase} we present the disorder phase diagram.
Next, in section \ref{Sec:Correlated} we repeat the analysis for the case of noise with a non-flat power spectrum (which is correlated along the length of the system).
Section \ref{Sec:Spectrum} is devoted to the power spectra of the response. Namely we establish a fairly universal spectral
response for the charge density and the condensate as a function of the spectral description of the disordered chemical
potential. We conclude in section \ref{Sec:Conclusions} where we also point out some interesting directions.

\section{Review of the holographic p-wave superconductor}\label{Sec:ReviewPW}

To build a holographic $p$-wave superconductor in 2+1
dimensions we start with the action introduced originally in \cite{Gubser:2008zu} and further studied in
\cite{Gubser:2008wv}. Namely, we consider the dynamics of a $SU(2)$
Yang-Mills field in a gravitational background:
\be S=\int d^4
x\,\sqrt{-g}\left(\frac{1}{16\pi G_N}\left(R-\Lambda\right)-{1\over4 q^2}Tr\,F_{\mu\nu}\,F^{\mu\nu} \right)\,.
\label{action}
\ee
In the limit where $G_N/q^2$ is very small, gravity can be considered decoupled and then, the Yang-Mills system is studied on the  Schwarzschild-AdS
metric:
\bea
ds^2&=&{1\over z^2}\left(-f(z)dt^2+{dz^2\over
f(z)}+dx^2+dy^2
\right),\nonumber \\
f(z)&=&1-z^3\,,
\eea
where we have set the radius of AdS, $R=1$, and
the position of the horizon to $z_h=1$.

In \cite{Gubser:2008wv} an Ansatz was chosen such that the spatial
rotational symmetry is spontaneously broken when the condensate
breaking the gauge $U(1)$ symmetry subgroup arises at low temperatures.

The field strength in the action Eq. (\ref{action}) is given by
\be
F^a_{\mu\nu}=\partial_\mu A^a_\nu -\partial_\nu A^a_\mu +f^a_{\,\,bc}A^b_\mu A^c_\nu\,,
\ee
and the corresponding Yang-Mills equations of motion:

\be
\nabla_\mu F^{a \mu\nu}+f^a_{\,\,bc}A^b_\mu F^{c\mu\nu}=0\,.
\ee
In what follows we specialize to $SU(2)$ with the following generators
(further conventions are as in \cite{Amado:2013xya}):
\be
\label{generators}
T_i=\frac{1}{2}\s_i\,,\hspace{1cm} \lbrace T_i,T_j \rbrace
=\frac{1}{2}\delta_{ij}\mathbb{I}\,.
\ee

It is possible to consider more general groups, for example $U(2)$, see \cite{Amado:2013aea,Amado:2013lia}.

Motivated by condensed matter applications and, in particular  superconductivity,
we are interested in a system at finite chemical potential which develops an instability at low temperatures.
One simple Ansatz that achieves this goal is
\be
A=\phi(z)dt\,T_3+w_x(z) dx \, T_1\,.
\ee
Following the AdS/CFT dictionary, one reads field theory information from the boundary data of the gravity fields.
Namely, the boundary ($z\to 0$) values of the fields are:
\bea
\label{homogeneous}
&&\phi(z)=\mu -\rho\,z+o(z^2)\,, \nonumber \\
&&w_x(z)=w_x^{(0)}+w_x^{(1)}\,z+o(z^2)\,.
\eea
The field theory interpretation in terms of these boundary values is as follows: $\mu$ is a chemical potential, $\rho$ is the charge
density, $w^{(0)}_x$ is the source and $w^{(1)}_x$ is the vacuum expectation value of the vector order parameter.
Since we are interested
in  spontaneous symmetry breaking we will require the source to vanish $w^{(0)}_x=0$.
Notice that due to the rescaling that allowed us to set the horizon radius $z_h=1$, the chemical potential $\mu$ is actually
dimensionless and proportional to the ratio of the physical chemical potential to the temperature.
If one works in the grand canonical ensemble, where the chemical potential is held fixed, the temperature of the system is
thus given by $T\propto 1/\mu$. Hence in the rest of the paper we will only talk about $\mu$, with the understanding that it is
equivalent to the inverse of the temperature of the boundary field theory at fixed chemical potential.

An intuitive way of understanding the mechanism of condensation is as follows.
The gravity mode $w_x(z)$ has an effective mass of the form:
\be
m_{eff}^2= q^2 \,\,g^{tt}\,\,\phi^2 \,.
\ee

Since $g^{tt}<0$, as we increase the value of $\mu$ the effective mass decreases and goes below the  BF bound in a sufficiently
large region of space and, consequently, a zero mode of $w_x(z)$ develops at some $\mu_c$. Increasing $\mu$ above the
critical value, $\mu_c$, leads to the field condensing, and a new branch of solutions with nonzero condensate emerges.
This instability mechanism is fairly universal, appearing both in the $s$-wave \cite{Hartnoll:2008vx,Hartnoll:2008kx} and $p$-wave
\cite{Gubser:2008zu,Gubser:2008wv} holographic superconductors.


The asymptotic value of $\mu$ plays the role of chemical potential in the dual field theory. It is
worth mentioning that since the order parameter is determined by the asymptotic value of the field $w$ which is a vector,
we have a vectorial order parameter.

To complete the analogy with the superconducting phase transition
 the conductivities were computed for this system \cite{Gubser:2008zu,Gubser:2008wv}, and qualitative agreement
 was established.
More recent studies of this system, some taking into account the gravitational back-reaction,
include: \cite{Herzog:2014tpa,Basu:2009vv,Ammon:2009xh,Gangopadhyay:2012gx,Roychowdhury:2013aua,Arias:2012py}.

\section{ Holographic p-wave superconductor with disorder}\label{Sec:Disorder}

The main goal in this manuscript is the introduction of disorder in the $x$-direction of the field theory dual.
To be consistent with the
equations of motion, we are not allowed to choose the direction of the condensate freely as done in the previous section.
We, therefore,  consider the following consistent Ansatz for the matter fields:
\bea
\label{Ansatz}
A=\phi(x,z)\,
dt\,T_3+w_x(x,z)\,T_1\, dx+ w_y(x,z)\,T_1\,dy + \theta(x,z)\,T_2\,dt
\,,
\eea
where $T_i$ are the $SU(2)$ generators presented before in Eq. (\ref{generators}).

 The Yang-Mills equations of motion following from the Lagrangian (\ref{action}) and the Ansatz (\ref{Ansatz}) are:
\bea
 &&-\partial_z^2\phi-\frac{1}{f}\,\partial_x^2\phi+\frac1 f (w_x^2+w_y^2)\phi+
 \frac1  f \theta \partial_x w_x+
 \frac2  f   w_x\partial_x\theta  =0\,,\label{eomphi}\\
 &&\partial_z^2w_x+\frac{f'}{f}\partial_z w_x+\frac1{f^2}\left(\phi^2+\theta^2\right)w_x
 +\frac1{f^2} \phi \partial_x\theta -\frac1{f^2} \theta\partial_x\phi =0\,,\label{eomwx}  \\
 &&\partial_z^2\theta+\frac1f\partial_x^2\theta-\frac1f (w_x^2+w_y^2) \theta
 +\frac1{f}\phi\partial_x w_x+\frac2{f}w_x\partial_x\phi =0\,,\label{eomtheta}  \\
 &&\partial_z^2 w_y+\frac1f\partial_x^2 w_y+\frac{f'}{f}\partial_z w_y+
 \frac1{f^2}\left(\phi^2+\theta^2\right)w_y=0\,.\label{eomwy}   \eea 

These equations of motion satisfy the constraint
\be
\phi \partial_z
\theta-\theta\partial_z\phi-f\,\partial_z\partial_xw_x=0\,,
\ee
which is a  consequence of gauge fixing: $A_z=0$.

{

As in the previous case, discussed around Eqs. (\ref{homogeneous}), to uncover the physics of the dual field theory
we need to examine the boundary values of the supergravity fields. The near boundary asymptotics of the solutions to equations (\ref{eomphi}-\ref{eomwy}) is given, at small values of $z$,  by:
\bea
&&\phi(x,z)=\mu(x)-\rho(x)\,z+o(z^2)\,,\\
&&w_x(x,z)=w_x^{(0)}(x)+w_x^{(1)}(x)\,z+o(z^2)\,,\\
&&\theta(x,z)=\mu_2(x)-\rho_2(x)\,z+o(z^2)\,,\\
&&w_y(x,z)=w_y^{(0)}(x)+w_y^{(1)}(x)\,z+o(z^2)\,.
\eea
The values  $\mu(x)$ and $\rho(x)$ correspond to space-dependent
chemical potential and charge density, respectively. Turning on a
chemical potential in the direction $T_3$ means breaking
$SU(2)\rightarrow U(1)_3$. The functions $w_i^{(0)}(x)$ and
$w_i^{(1)}(x)$ are identified, under the holographic duality, with the source
and VEV of vectorial operators in the $i$ direction. Finally, $\mu_2(x)$ and
$\rho_2(x)$ are, respectively, a new chemical potential and charge density that
are sourced by the space-dependent condensate.

The near horizon conditions on the gravity fields are completely determined by regularity of the solution. Regularity,
consequently, implies that $A_t$ vanishes at the horizon. Hence, we consider an asymptotic expansion
about $z\sim1$ of the form

\bea
 &&\phi(x,z)=(1-z)\,\phi_h^{(1)}(x)+(1-z)^2\,\phi_h^{(2)}(x)+\ldots \,,\nonumber \label{phiIRexp}  \\
&& \phi(x,z)=(1-z)\,\theta_h^{(1)}(x)+(1-z)^2\,\theta_h^{(2)}(x)+\ldots \,,\nonumber \label{thetaIRexp}  \\
 &&w_i(x,z)=w_{ih}^{(0)}(x)+(1-z)\,w_{ih}^{(1)}(x)+(1-z)^2\,w_{ih}^{(2)}(x)+\ldots\,, \label{wIRexp}
\eea
where the ellipses stand for higher order terms.


For numerical reasons, we find it convenient to redefine some of the fields involved in the equations of motion. Namely:
\be
\chi_i(x,z) = (1-z)\,w_i(x,z)\,.
\label{chidef}
\ee

In terms of the redefined fields (\ref{chidef}) the equations (\ref{eomphi}-\ref{eomwy}) take the form
\begin{subequations}
\bea
 &&\partial_z^2\phi+\frac{1}{f}\,\partial_x^2\,\phi
 -{\chi^2\over(1-z)^2 f}\phi
 -{1\over(1-z)\,f}\left(2\chi_x\,\partial_x\theta+(\partial_x\chi_x)\,\theta \right)=0\,,\label{eomrdphi}\\
 &&\partial_z^2\theta+\frac1f\partial_x^2\theta
 -{\chi^2\over f(1-z)^2}  \theta
 +{2\chi_x\,\partial_x\phi+(\partial_x\chi_x)\,\phi\over f(1-z)}=0\,,\label{eomrdtheta}  \\
&&\partial_z^2\chi_x+\left(\frac{f'}{f}+\frac2{1-z}\right)\partial_z\chi_x
+{2f^2+(1-z)\,f\,f'+(1-z)^2(\theta^2+\phi^2))\over(1-z)^2\,f^2}\chi_x+\nonumber \\
&&+{1-z\over f^2}\left(\phi\,\partial_x\theta-(\partial_x\phi)\,\theta\right)
=0\,,\label{eomrdwx}  \\
 &&\partial_z^2 \chi_y+\frac1f\partial_x^2 \chi_y+\left(\frac{f'}{f}+\frac2{1-z}\right)\partial_z
 \chi_y+
 \left[
 {2\over (1-z)^2}+{1\over f^2}(\theta^2+\phi^2)+{f'\over f\,(1-z)}
 \right]\chi_y=0\,,\label{eomrdwy}\nonumber\\   \eea
 \label{pderw}
 \end{subequations}
where
$\chi^2=\chi_x^2+\chi_y^2$. 

This redefinition leads to a simpler set of boundary conditions:
\bea
\chi_i(x,0)&=&0,\qquad  \theta(x,0)=0\,, \qquad  \phi(x,0)=\mu(x)\,,\; {\rm UV}\,\,\,  z\to 0\,, \nonumber \\
\chi_i(x,1)&=&0\;,\qquad \theta(x,1)=0\,,\qquad
\phi(x,1)=0\,,\;\;\;\;\; {\rm IR}\,\,\, z\to 1\,.\label{bcirpsi}
\eea

This choice of boundary conditions corresponds to a spontaneous
breaking of the $U(1)$ symmetry with order parameter $\langle {\cal
O}\rangle\propto w^{(1)}_i(x)$. From now on we use the angle brackets
associated with ${\cal O}$ exclusively to refer to the average over $x$.
Moreover, we choose the condition $\mu_2=0$ (vanishing source for the charge
density in the $T_2$ direction) since we want a disordered
version of the p-wave superconductor. Hence, the charge density $\rho_2$
will be spontaneously induced. In this case, since the symmetry of our
action (\ref{action}) is the whole $SU(2)$, we are effectively realizing a two-component superfluid as recently discussed in the holographic framework in \cite{Amado:2013xya} following original ideas of  \cite{PhysRevB.11.178},
an important
phenomenological paradigm in various condensed matter situations. We will however not pursue these questions in the present manuscript.

It is worth noticing that there are two simplified situations
that might be taken into account, given that they might include all the
interesting physics but require less computing power. It is easy to
see that the equations (\ref{eomphi}-\ref{eomwy}) allow to consistently set
$w_x=\theta=0$ or $w_y=0$. This is equivalent to going to the two
limits in which the system condenses in the direction of the noise
or in the direction perpendicular to the noise. We will devote ample attention to these branches in section \ref{Sec:Thermo}.

\subsection{Introducing disorder}
\label{Subsec:disorder}
We are interested in solving the system given by equations (\ref{pderw}) 
in the presence of disorder. Let us take the following form for the noisy chemical potential:
\bea
\nonumber
\mu(x)&=&\mu_0+\epsilon\sum_{k=k_0}^{k_*}{\sqrt{S_k}}\,\cos(k\,x+\delta_k)\\
&=&\mu_0+\epsilon\sum_{k=k_0}^{k_*}{1\over
k^\alpha}\,\cos(k\,x+\delta_k)\,, \label{noisefunc}
\eea
where $\delta_k$ is a random phase for each $k$ and $S_k$ is the power
spectrum.
For the case  $\alpha=0$, corresponding to a flat spectrum, in the limit of infinitely many modes ($k_*\to\infty$)
the function in Eq. (\ref{noisefunc}) tends to a Gaussian distributed random function.
Instead, as we will see below, for $\alpha>0$, the typical length scale of the noise is of the order of the system size.
In this latter case, the power of $1/k$ determines the differentiability properties of $\mu(x)$.

Let us now discuss, in detail, the properties of the correlation function of the disorder we introduce above.
It takes the form
\be
\left< \mu(x)\mu(0) \right>-\mu_0^2=\sum_{k=k_0}^{k_*}\frac1{k^{2\alpha}}\cos(kx)\,.
\label{corrfunc}
\ee
For $\alpha>0$ this correlation function is periodic and does not diverge for large $x$.
%
%
Moreover, one could try and go to the continuum limit, replacing the sum in Eq. (\ref{corrfunc}) by an integral over $k$. The corresponding expression for the correlation function is then given by
\be 
\left< \mu(x)\mu(0) \right>-\mu_0^2\approx |x|^{2
\alpha-1}\Gamma(1-2\alpha)\sin(\alpha\, \pi)+
\frac{k_0^{1-2\alpha}}{2\alpha-1}   \,
_1F_2\left(\frac12-\alpha;\,\frac12,\, \frac32-\alpha ;\, -\frac14
k_0^2 x^2\right)\,. 
\ee
It is clear from this expression that thanks to the IR cutoff $k_0$, no divergence appears in the correlation function.
Note, however, that the first term grows with distance, the second term kills that divergence and the result is a damped oscillatory correlation function.
More explicitly, the large $x$ limit of the expression above reads
\be
\left< \mu(x)\mu(0) \right>-\mu_0^2\rightarrow \frac{k_0^{-2
\alpha}}{x}\cos(k_0 x)+\dots\,, 
\label{xilargex}
\ee
The power law decay of the correlation function, reminiscent of a critical system, does not allow us to define a correlation length by the usual prescription. Lacking this, the oscillations define a length scale $\propto 1/k_0$, which could be as large as the system size.
Therefore, in a broad sense we can say that for $\alpha>0$ our noise is correlated along the whole system; we will denote this case as correlated noise.

Low correlation in the noise can be achieved by taking $\alpha=0$ in (\ref{noisefunc}), which results in a correlation function of the form
\be
\left< \mu(x)\mu(0) \right>-\mu_0^2=\sum_{k=k_0}^{k_*}\cos(k x)
={\rm Re}\left(
e^{ik_0\,x}\,{e^{i(k_*-k_0+1)\,x}-1\over e^{ix}-1}
\right)\,.
\label{correlength}
\ee
Notice that this function, which we plot in Fig. \ref{corrfig},
is the closest to a delta function one can get with a finite number of modes. 
Hence, in order to study a more realistic realization of noise, in the bulk of the work presented in this article we consider the case where $\alpha=0$, which we will denote as uncorrelated noise.
However, the correlated noise ($\alpha>0$) presents interesting features, such as a well defined continuum limit.
Moreover, as in \cite{Arean:2013mta}, it will allow us to study the power spectra of the response functions in our setup. Therefore, in sections \ref{Sec:Correlated} and \ref{Sec:Spectrum} we will analyze the case of correlated noise
($\alpha>0$).



To implement the noise given by Eq. (\ref{noisefunc}) and solve the coupled PDEs (\ref{pderw}),
we discretize the space and impose periodic boundary conditions in the $x$ direction, leading to a discretized
$k$ with values:
\be\label{Eq:k-range}
k_n={2\pi\,n\over L}\quad {\rm with}\quad 1\leq n\leq{L\over 2a_x}\,,
\ee where $L$ is the length in the $x$ direction of our cylindrical
space, and $a_x$ is the lattice spacing in $x$. Note that there is an
IR scale given by $k_0$ and a UV scale defined by $k_*$ \footnote{Notice that both these two scales, as well
as the chemical potential $\mu$ are measured in terms of the temperature, since we have made use of the scaling symmetries of the problem to fix the horizon of the black hole at $z_h=1$.}. 
Notice that the UV scale $k_*$ is given by $L/(2a_x)$, and
was chosen here to saturate the Nyquist limit\footnote{Nyquist frequency is the highest frequency that can be reconstructed from a signal given a sample rate. In order to recover all Fourier components of a periodic waveform, it is necessary to use a sampling rate at least twice the highest waveform frequency.
This can be understood from the fact that there are two Fourier coefficients to fit for each frequency.}. 
However, as we explain below, when performing our numerical simulations we will take a $k_*$ sensibly smaller than
$L/(2a_x)$, in order to allow for our lattice to be sensible to higher harmonics sourced by our noise.
In order to parametrize the strength of the noise,  which in Eq. (\ref{noisefunc}) is characterized by $\epsilon$, let
us introduce the variable $w$ defined through the expression
\be
w={25\epsilon\over\mu_0}\,,
\label{wdef}
\ee
so that $w$ corresponds to a strength relative to the chemical potential $\mu_0$\footnote{The factor ${1\over25}$
is included to keep the same normalization as in \cite{Arean:2013mta}}. 
Naturally, $w=0$ corresponds to the homogeneous case, while 
the largest $w$ will be chosen by demanding that $\mu(x)$  remains positive all along the system.
Notice that this maximum value of $w$ will depend on the scales $k_0$ and $k_*$, and for the case of correlated noise also on the power $\alpha$ characterizing the power spectrum.
Our definition of $w$ corresponds, in the standard solid
state notation, to $1/k_F\, l$, where $k_F$ is the Fermi momentum and $l$ is the mean free path \cite{phillips2012advanced}.


\subsubsection*{Numerical Methods}
In order to solve the system of PDEs we have discretized it on a rectangular lattice of size $N_z\times N_x$, where $N_z$
and $N_x$ correspond, respectively, to the number of points in the $z$ and $x$ directions.
We used planar lattices for the $x$ direction, and Chebyshev grids along $z$.
Consequently, the
discretization of the derivatives was performed using pseudo spectral methods (with periodic boundary conditions
in the $x$ direction). To find solutions we employed a Newton-Raphson algorithm on lattices with a typical
size of $25\times 90$. 

\begin{figure}[htp]
\begin{center}
\includegraphics[width=3.5in]{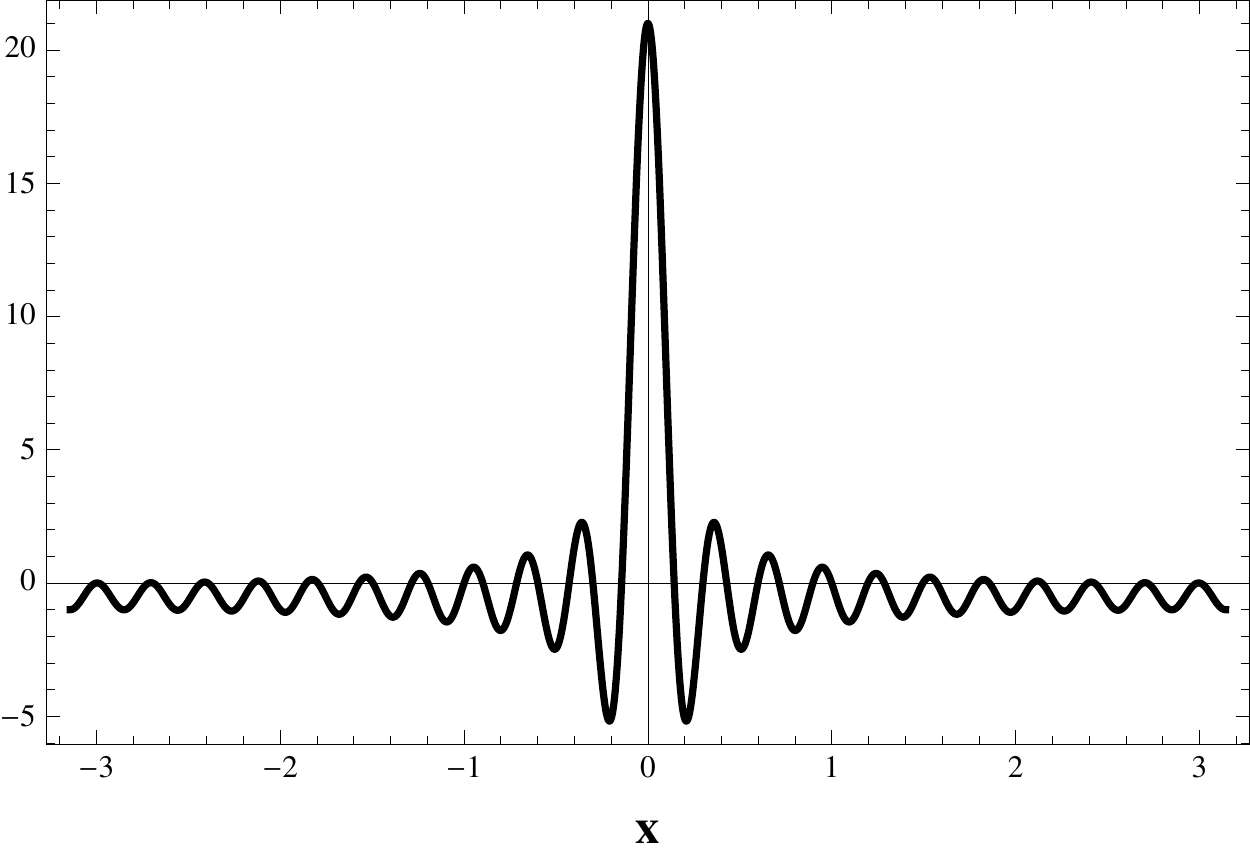}
\caption{\label{corrfig} Plot of the correlation function of our chemical potential, Eq. (\ref{correlength}),  for the parameters used in our numerical simulations: $L=2\pi$, $k_*=21$.}
\end{center}
\end{figure}

In the simulations performed to determine the phase diagram of our setup we set the system length to $L=2\pi$ and take $k_*=21$ and $\alpha=0$. We are therefore truncating the sum in Eq. (\ref{noisefunc}) at 21 modes. Notice that for a lattice with 90 points along the $x$ directions, saturating the Nyquist limit would correspond to a $k_*=45$, and hence 45 modes in the sum (\ref{noisefunc}).
For this parameters, from Eq. (\ref{correlength}) we can establish a correlation length 
about $2.3\%$ of $L$ (the system size) or 2 lattice spacings.
Finally, in Fig. \ref{corrfig} we plot the correlation function (\ref{correlength}) for this choice of parameters.
%
%
%

%
%

\vspace{.5cm}

In Figs.  \ref{ModelSimulationa0}, \ref{ModelSimulationa02} we plot the results of a single typical simulation. In  Fig. \ref{ModelSimulationa0} we present the random chemical potential $\mu(x)$ (upper left panel) together with the solutions
for the fields $\phi$, $\theta$, $\chi_x$ resulting from solving the system (\ref{pderw}) with that chemical potential as boundary condition.
Instead, in Fig.  \ref{ModelSimulationa02} we present the boundary data read from this same solution.
First, notice that the plots presented in these figures correspond to a solution for which the condensate lies purely in the $x$ component of the vector parameter. We will comment on
the competition between different branches of solutions (with condensate parallel or perpendicular to the direction of the noise) in section \ref{Sec:Thermo}.

As expected, the introduction of disorder  leads to 
a space-dependent charge density and condensate in some cases which we plot in those figures.
In a sense one could view the gravity equations of motion as a tool that provides precise answers to the question: Given a random chemical potential in a strongly coupled system with a superconducting transition, what is the value of the condensate and the charge density that the system uses to respond to the random chemical potential?
\begin{figure}[tb]
\begin{center}
\begin{subfigure}[b]{0.40\textwidth}
\includegraphics[width=\textwidth]{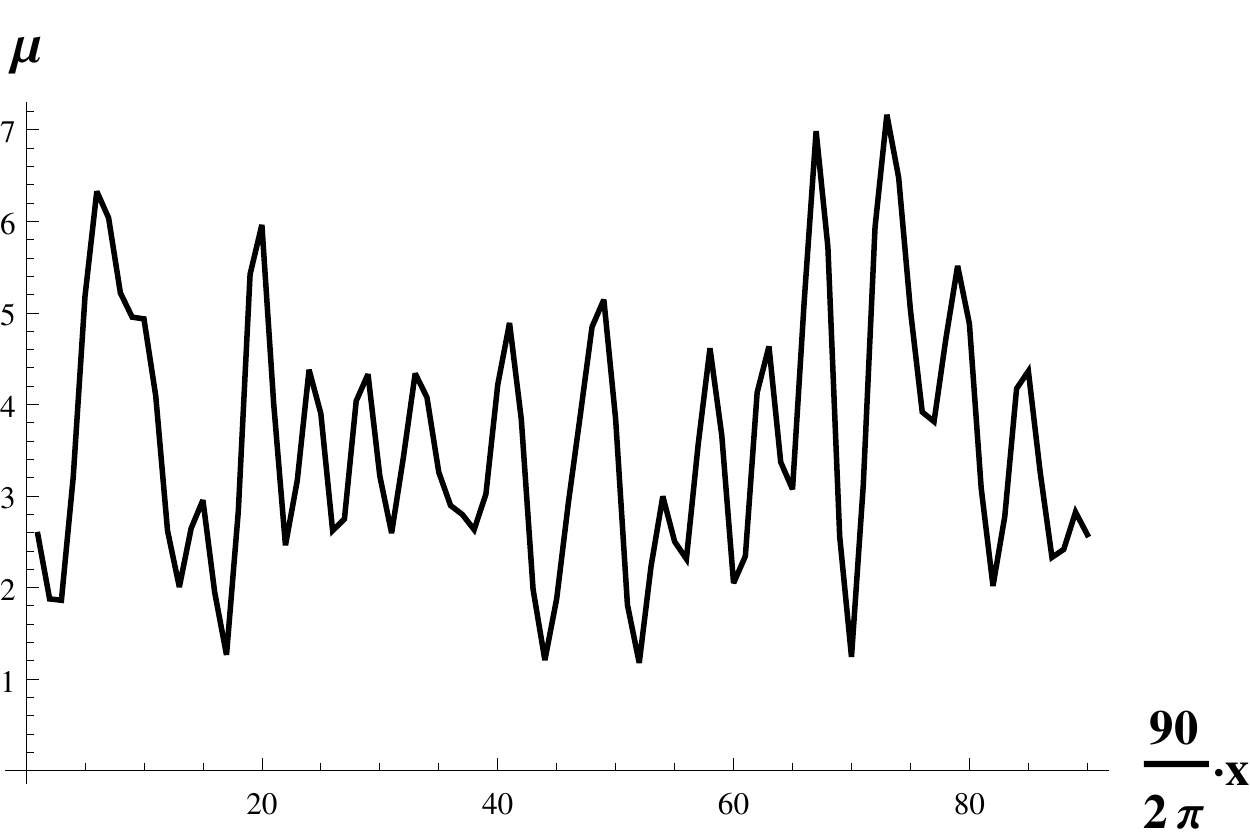}
\end{subfigure}
~
\begin{subfigure}[b]{0.40\textwidth}
\includegraphics[width=\textwidth]{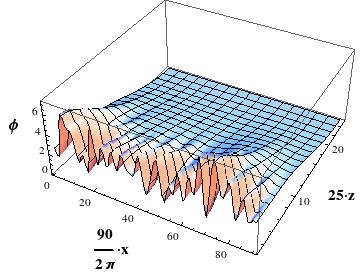}
\end{subfigure}\\[4mm]

\begin{subfigure}[b]{0.40\textwidth}
\includegraphics[width=\textwidth]{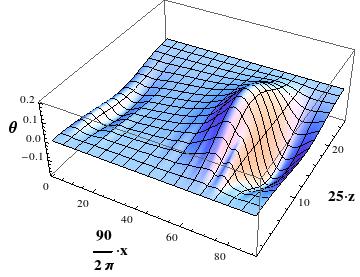}
\end{subfigure}
~
\begin{subfigure}[b]{0.40\textwidth}
\includegraphics[width=\textwidth]{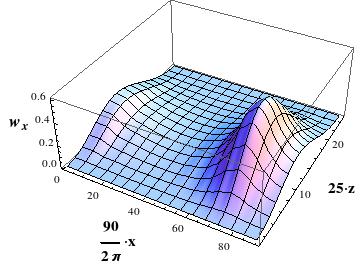}
\end{subfigure}
\caption{Example of a simulation corresponding to a noisy chemical potential with
$w=2.95$, $\alpha = 0$, and $\mu_0=3.55$ below the critical value of the homogeneous chemical potential
($\mu_c=3.66$). We plot the original spatial-dependent chemical potential on the upper-left panel and the
corresponding solutions for the fields $\phi$, $\theta$ and $\chi_x$ ($\chi_y=0$ for this solution)
on the other three panels.} \label{ModelSimulationa0}
\end{center}
\end{figure}

\begin{figure}[tb]
\begin{center}
\begin{subfigure}[b]{0.40\textwidth}
\includegraphics[width=\textwidth]{sim2da0_mu.pdf}
\end{subfigure}
~
\begin{subfigure}[b]{0.40\textwidth}
\includegraphics[width=\textwidth]{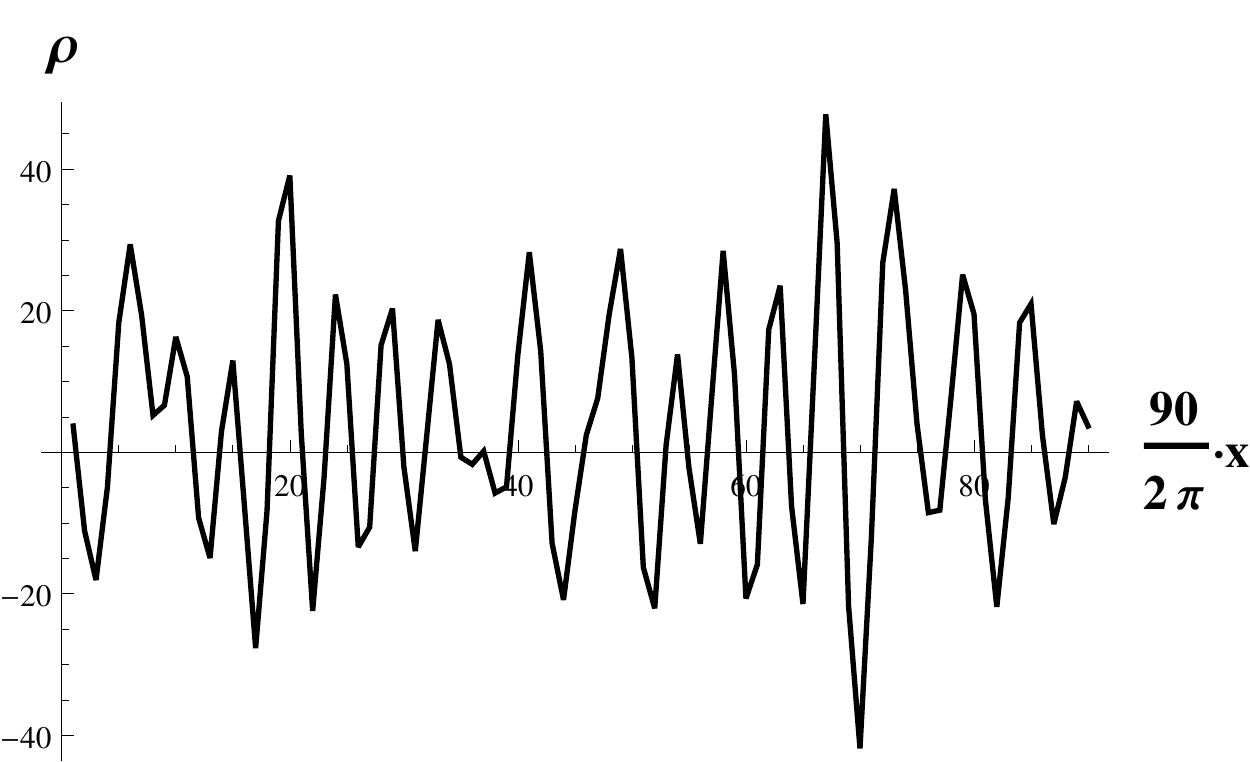}
\end{subfigure}\\[4mm]

\begin{subfigure}[b]{0.40\textwidth}
\includegraphics[width=\textwidth]{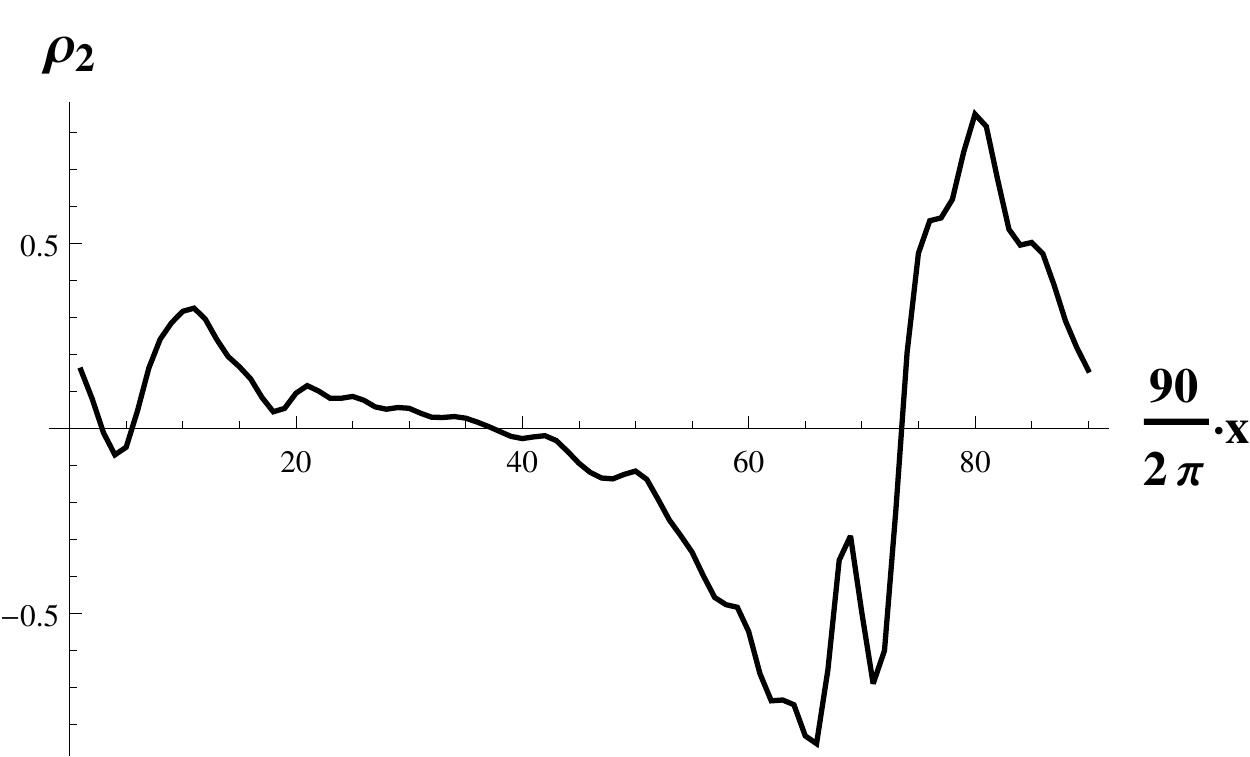}
\end{subfigure}
~
\begin{subfigure}[b]{0.40\textwidth}
\includegraphics[width=\textwidth]{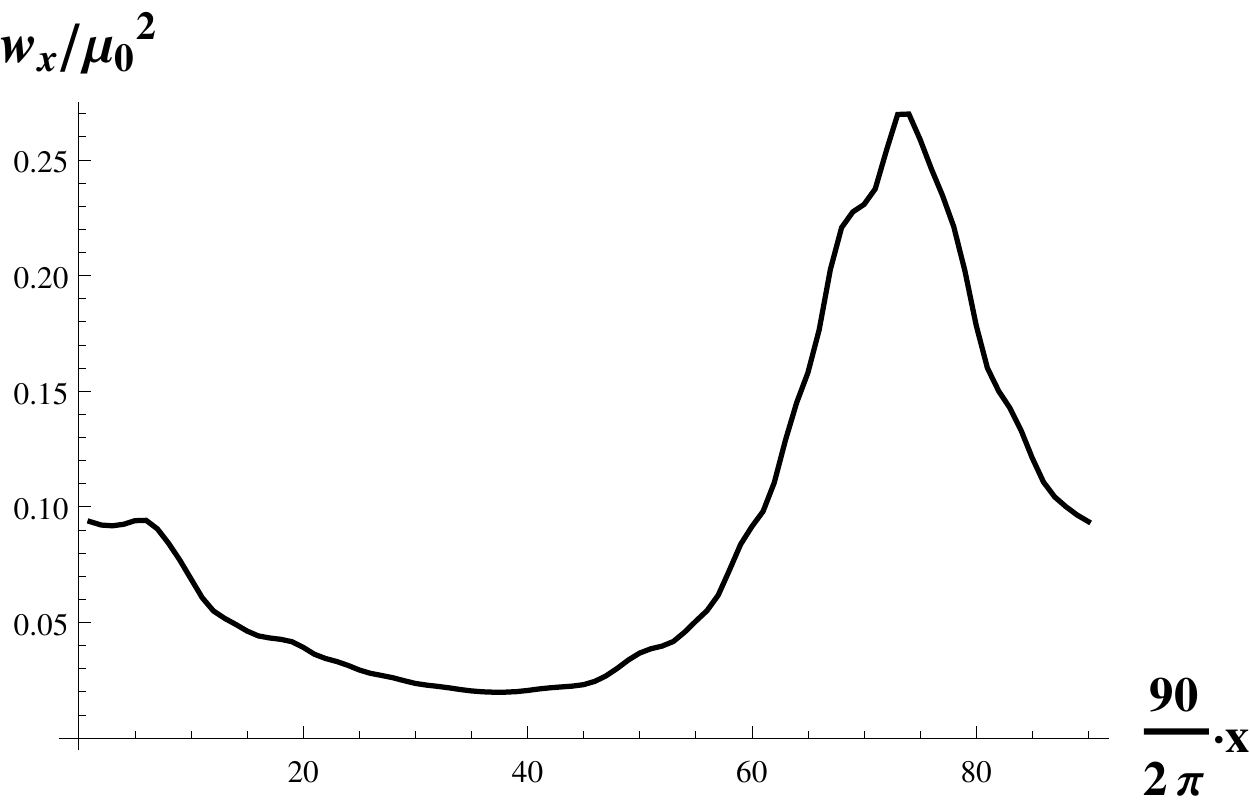}
\end{subfigure}
\caption{Boundary data corresponding to the same simulation as in Fig.
\ref{ModelSimulationa0}. From the upper left panel and in clockwise sense:
$\mu(x)$, $\rho(x)$,  $w_x(x)$, and $\rho_2(x)$.} \label{ModelSimulationa02}
\end{center}
\end{figure}


%
%

Notice that the response to the source $\mu(x)$ is noisier for the VEV corresponding to the same field, that is, $\rho(x)$; while the VEVs realized by
other fields, namely $w_x(x)$ and $\rho_2(x)$, are visibly smoother. We will investigate this behavior in more detail in section \ref{Sec:Spectrum}.

It is worth pointing out, and we will use this result in the upcoming sections, that we find that for a chemical potential below the critical (in the homogeneous case), there are values of the strength of the disorder that render the expectation value everywhere non-vanishing; this is our definition of arriving at the superconducting phase via disorder.

\subsection{Thermodynamic limit and self-averaging condensate}

In the context of condensed matter physics it is quite important to consider the thermodynamic limit.
Namely, to study the properties of the system in the limit where its size goes to infinity.
The thermodynamic limit plays a particularly important role in systems dominated by quenched randomness.
In this subsection we address some issues related to the thermodynamic limit as they apply to the the problem at hand.

Let us review, once more, all the scales involved in the problem, both, physical and numerical.
The three physical dimensionless scales of the problem are $T/\mu$, $w$, and $L\,mu$ corresponding respectively to
the temperature, the strength of the noise, and the length of the system in the $x$ direction, all measured relative to
the chemical potential.
Since we are solving the problem numerically using a lattice we have two extra scales: $a_x$ and $a_z$ which are the sizes
of cells along the $x$ and the $z$ directions.

Discussing the thermodynamic limit is more than a mere academic question.
Given that the realization of disorder is intrinsically related to our way of solving the system we need to show that there
is a limit to which we are truly approximating. In general, we expect that in the thermodynamic limit certain quantities
will be self-averaging. A property $X$ is self-averaging if most realizations of the randomness have the same value of $X$.
More precisely, in the numerical context we use that: The system is said to be self-averaging with respect to property $X$ if
\be
\frac{<X_n^2>-<X_n>^2}{<X_n>^2} \to 0,
\ee
as the size, $n$, of the system goes to infinity.
Here the angular brackets denote averages over the realizations of the quenched randomness of the system and $X_n$ is the value
of property $X$ when the system has size $n$  \cite{0305-4470-35-19-303,Binder:1986zz}.

\begin{figure}[tp]
\begin{center}
\includegraphics[width=6.5in]{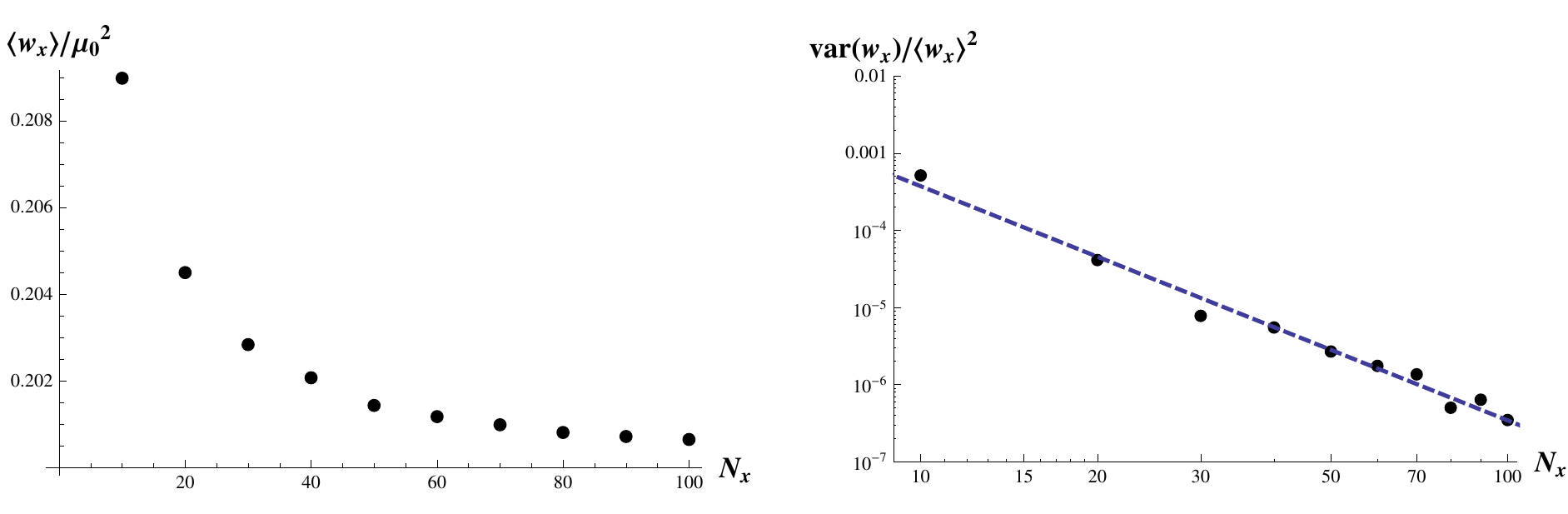}
\caption{\label{Fig:LogLog} On the left panel we plot the average of the condensate versus the number of lattice
sites $N_x$, notice that it stabilizes in the thermodynamic limit.
The right panel shows the variance of the condensate versus $N_x$, and the blue dashed line shows the fit
$\log({\rm var}(w_x)/\langle w_x\rangle^2) =-0.90-3.03\,\log(N_x)$. These figures show that the condensate $w_x$ is self-averaging. The data results from averaging over 50 realizations of a noisy chemical potential with $\alpha=0$,
$\mu_0=4.20$, and $w$ such that $w^2=160/N_x$ (so that the variance of the noisy chemical potential is kept
constant as $N_x$ is increased). }
\end{center}
\end{figure}

We will define our thermodynamic limit as the limit in which the correlation length of the disorder is negligible with respect to the length of the system.
In order to do so we will work with flat spectrum noise, which corresponds to setting $\alpha=0$ in Eq.
(\ref{noisefunc}).
Then
the scale $a_x$ (wich sets $k_{*}=\pi/a_x$) will determine the correlation length
of the disorder in our lattice.
Increasing the number of points $N_x$ of the lattice in the $x$ direction while keeping
the length of the system fixed will now imply to decrease the correlation length of the noise with respect
to the size of the system.
Using this prescription, we now study the self-averaging property of the condensate in the $x$-direction, $w_x$.
We claim that the average of the condensate stabilizes, while its variance goes to zero. Moreover,
we provide numerical evidence that this vanishing goes as a power law  $\sim N_x^{-3}$ (see Log-Log plot in
Fig. \ref{Fig:LogLog}).

\begin{figure}[tp]
\begin{center}
\includegraphics[width=6.5in]{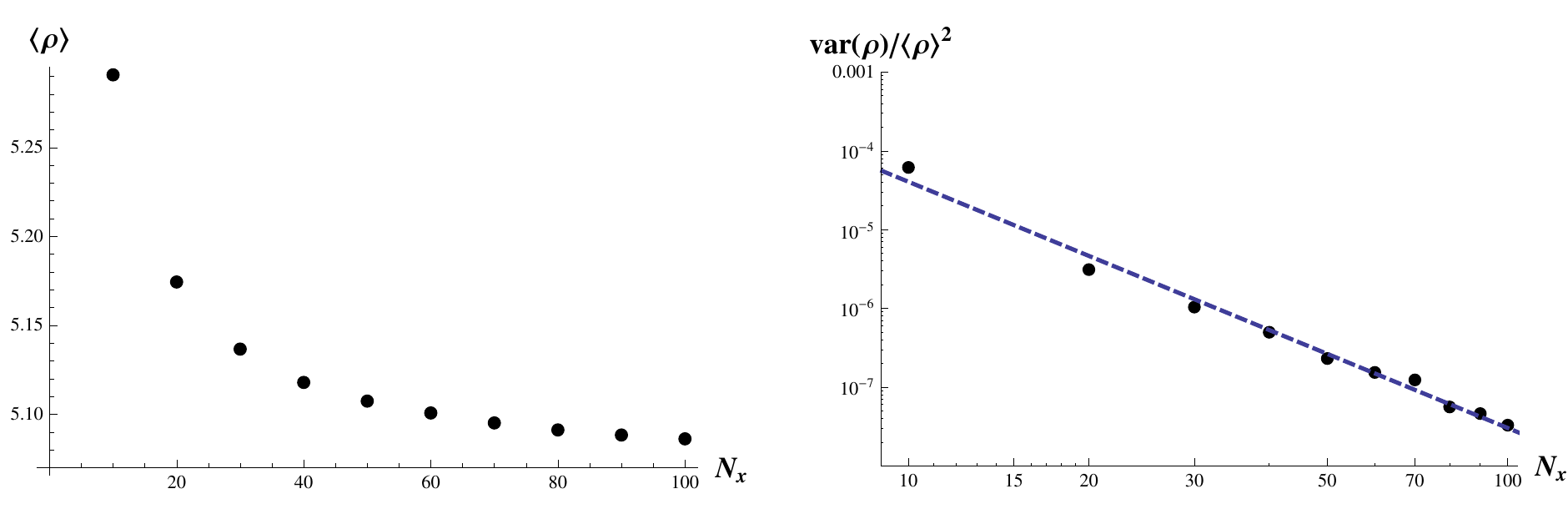}
\caption{\label{Fig:LogLogrho} On the left we show the average of the charge density $\rho$ versus the number of lattice sites $N_x$, while on the right we plot the variance of $\rho$ versus $N_x$. The blue dashed line results from the fit $\log({\rm var}(\rho)/\langle\rho\rangle^2) =-2.92-3.13\,\log(N_x)$. These plots follow from the same set of
data as those in Fig. \ref{Fig:LogLog}.}
\end{center}
\end{figure}

The same procedure as described above can be applied to study the self-averaging property of the charge density, $\rho$.
The result is presented in figure \ref{Fig:LogLogrho}, which clearly shows that the charge density is indeed self-averaging
in the thermodynamic limit.




\section{Free energy and competing solutions}\label{Sec:Thermo}

As already anticipated when we wrote the system of equations in section \ref{Sec:Disorder},
there are different branches or consistent truncations of the system of equations (\ref{pderw}).
One could expect three main types of solutions:

\begin{itemize}
 \item Solutions with $\chi_x\equiv 0$ and also $\theta\equiv 0$, we will denote these solutions by  $Y$, since the
 vector order parameter lies in the direction transverse to the noise.
 Note that in this approximation the system (\ref{pderw}) reduces significantly.
 In particular, equations (\ref{eomrdtheta}) and (\ref{eomrdwx}) are identically satisfied.

\item Solutions where $\chi_y\equiv 0$, we denote them by $X$, as they correspond to a vector condensate along the $x$
direction. In this limit the system of equations (\ref{pderw}) leads to a system with equation (\ref{eomrdwy})
trivially satisfied.

\item A third possibility would be that of solutions where all functions in the system (\ref{pderw})
are nonzero. These would correspond to a vector condensate pointing along an intermediate direction in the $x\,y$ plane.
However, our numerics indicate that these solutions do not exist.
\end{itemize}

Let us elaborate a bit more about the absence of solutions where the condensate lies along an intermediate direction in the
$x\,y$ plane. As is clear from the equations, in the absence of noise the system in the normal phase is rotational invariant.
Therefore, symmetry-breaking solutions with the condensate along any arbitrary direction on the plane are equivalent.
However, as soon as some noise is turned on, our numerics converge to solutions with the condensate being either parallel or
orthogonal to the noise. We checked this fact by starting from a broad family of seeds.

We will now study the free energy of the $X$ and $Y$ solutions to decide which of them is energetically favorable.
The free energy of the system is given in terms of the on-shell action (\ref{action}) as
\bea
\Omega&=&- \frac{T S_{\rm on-shell}}{ L_y\,L}= \\
&=&-\frac1{4L}\int_0^L dx \,\mu\, \rho+\frac1{4L}\int_0^L dx\int_0^1
dz\,\frac1f\left[(\theta^2+\phi^2)\,(w_x^2+w_y^2)+w_x (\phi\, \partial_x
\theta-\theta\,\partial_x\phi)\right]\,,\nonumber
\eea
where $L_y$ is the length of the system in the $y$ direction; this is a regulator
we need in order to get a finite result and will simply cancel out when integrating along the $y$ direction since
the solutions are $y$ independent.

In Fig. \ref{SvW} we plot the free energy for the two kinds of superconducting solutions, subtracted from
that of the normal phase\footnote{In the normal phase $\theta=\chi_x=\chi_y=0$, and the system (\ref{eomrdphi}-\ref{eomrdwy}) reduces to the equation (\ref{eomrdphi}). The normal phase solution exists for all values of $\mu$.}.
We observe that, when it exists, the $X$ solution has always lower free energy.
Therefore, in the rest of the paper when we refer to the superconducting phase we will restrict ourselves to the X
solutions, namely those with condensate pointing in the direction parallel to the noise.

\begin{figure}[htp]
\begin{center}
\includegraphics[width=3.0in]{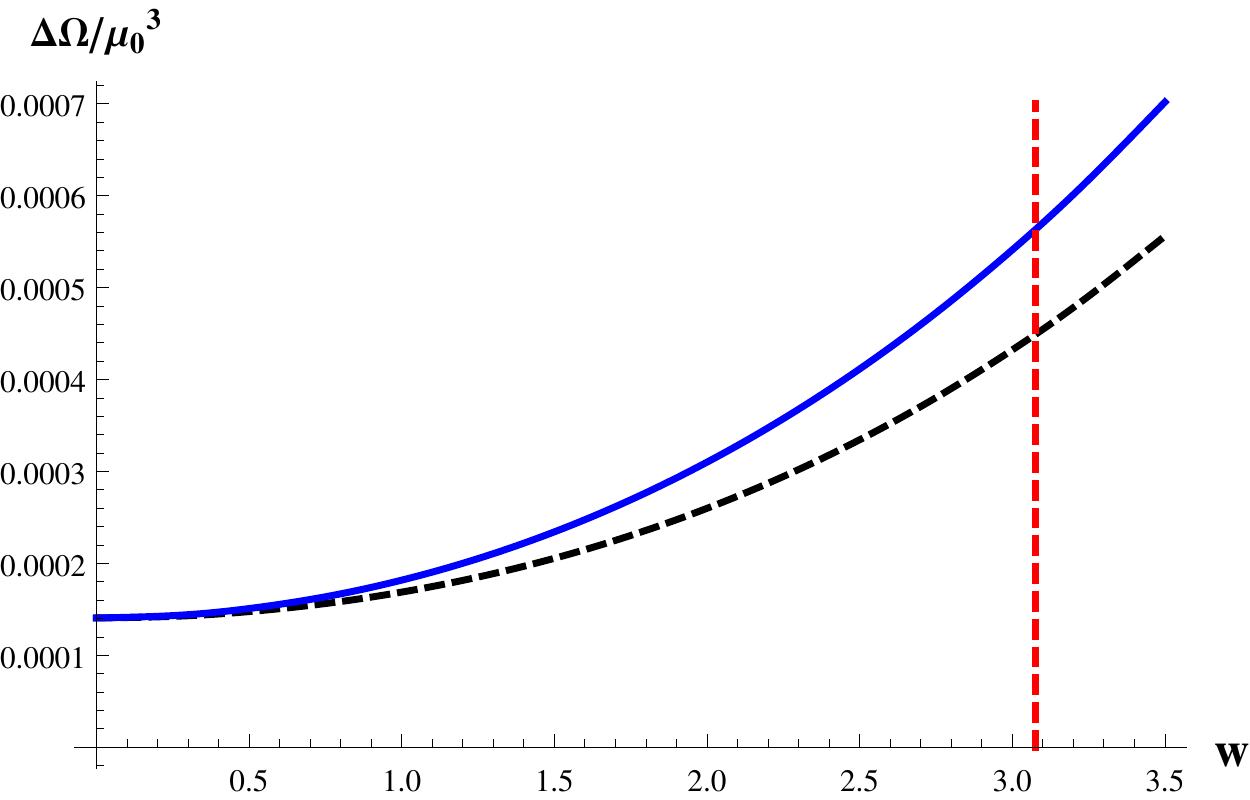}
\includegraphics[width=3.0in]{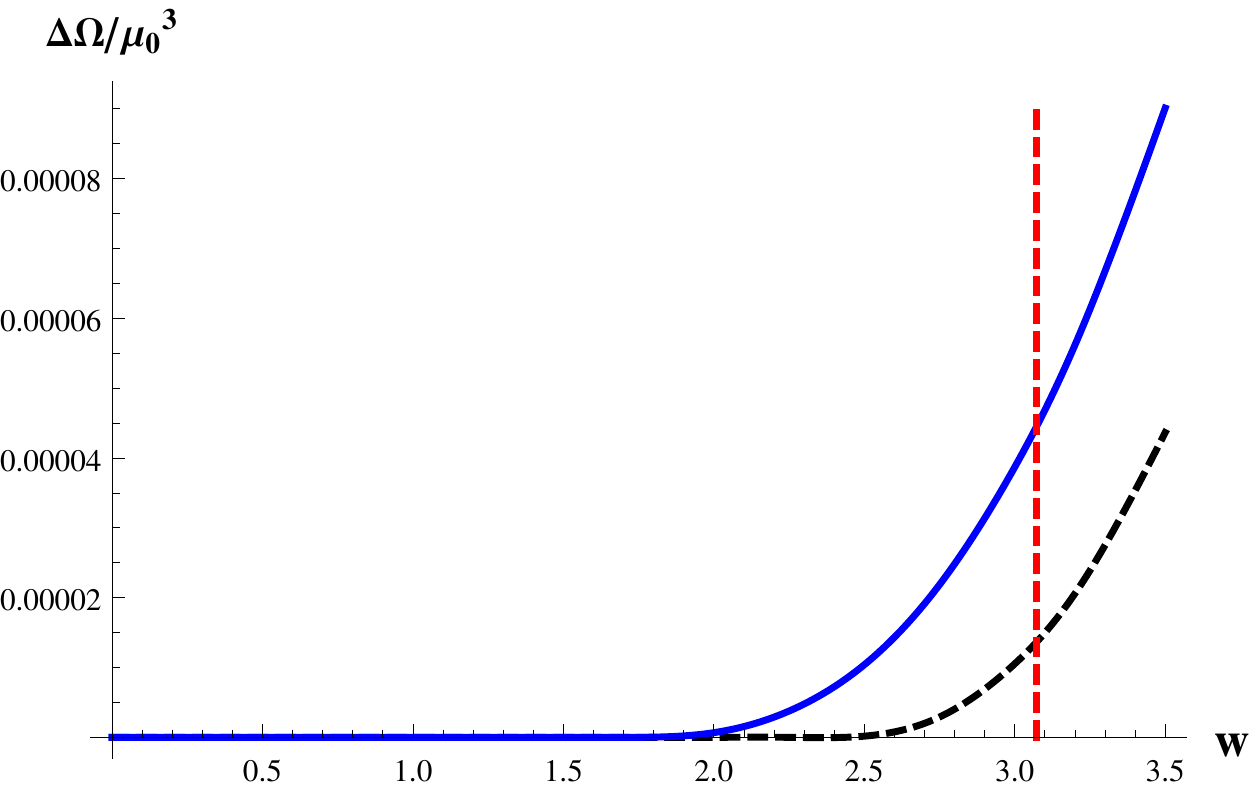}
\caption{\label{SvW} Subtracted free energy of competing solutions as a function of the strength of
the disorder,  $w$. The blue (black) solid (dashed) line corresponds to the free energy of the X
(Y) solution subtracted from that of the normal phase. On the left panel we present the results for an average
chemical potential $\mu_0=3.8$ above the critical one. The plot on the right corresponds to $\mu_0=3.55$ below
the critical $\mu_c=3.66$. These plots result from averaging over 5 realizations on lattices of
size $25\times90$.}
\end{center}
\end{figure}

We shall now provide a heuristic explanation for the fact that the $X$ solution
is energetically favorable (see \cite{Basu:2008st,Erdmenger:2013zaa} for similar arguments).
The key observation follows from comparing the equations for the fields $w_x$ and $w_y$, namely
Eqs. (\ref{eomwx}-\ref{eomwy}), which we reproduce for convenience
\bea
&&\partial_z^2w_x+\frac{f'}{f}\partial_z w_x+\frac1{f^2}\left(\phi^2+\theta^2\right)w_x
 +\frac1{f^2} \phi \partial_x\theta -\frac1{f^2} \theta\partial_x\phi =0\,, \\
 &&\partial_z^2 w_y+\frac{f'}{f}\partial_z w_y+
 \frac1{f^2}\left(\phi^2+\theta^2\right)w_y+\frac1f\partial_x^2 w_y=0\,,
 \eea
We can, for example, consider that the mode $w_y$ is governed by an effective mass of the form:
\be
m^2_{w_y}\propto -\frac1{f^2}\left(\phi^2+\theta^2\right)-\frac1{w_y f}\partial_x^2 w_y.
\ee
As noticed already in \cite{Erdmenger:2013zaa} the last term contributes a positive amount to the effective mass and
impedes condensation\footnote{Notice that at linear level one could Fourier transform the equation and consider the
effect of a single wave by replacing $\partial_x$ by $i\,k$.}.
Note, however, that the situation is different for the equation describing the effective mass for
the mode $w_x$.
\be
m^2_{w_x}\propto -\frac1{f^2}\left(\phi^2+\theta^2\right)-\frac1{w_x f^2} \left(\phi \partial_x\theta -
\theta\partial_x\phi\right)\,.\label{meffwx}
\ee
There is no term $\sim \partial_x^2 w_x$ impeding condensation, and it can be argued that the last term in
the above effective mass is small, since the derivatives cancel at linear level.
Thus, condensation of the mode $w_y$ seems to be disfavored while for the mode $w_x$ it is not.
Then, one may expect the free energy of the $X$ solution to be lower than that of the $Y$ solution,
as our numerics show.

Finally, let us speculate on the consequences the outcome of this free energy computation may have
for more realistic systems with bidimensional inhomogeneities.
Since it turns out that the solutions with condensate parallel to the noise are always thermodynamically preferred,
one would expect that in the presence of disorder in two spatial directions the condensate would point in the stiffest
direction, thus following the gradient of the bidimensional noise.

\section{Toward the disordered phase diagram}\label{Sec:Phase}

In this section we present the phase diagram of the disordered holographic p-wave superconductor.
The key strategy is to repeat the simulations outlined in section \ref{Sec:Disorder}  with a random chemical potential $\mu(x)$ given by Eq. (\ref{noisefunc}) in the regime illustrated by figure \ref{corrfig}, and to do that enough times so that we develop meaningful statistics.
\begin{figure}[htb]
\begin{center}
\includegraphics[width=0.95\textwidth]{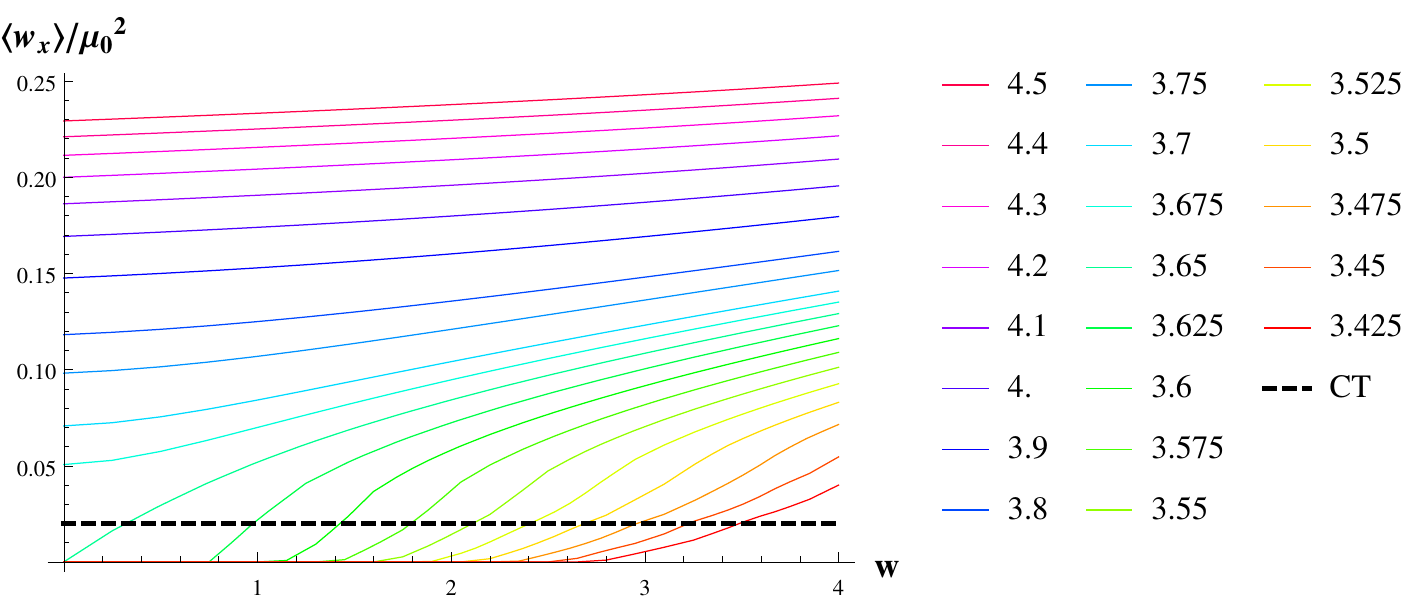}
\caption{\label{Cx-v-N} Spatial average of the condensate as a function of the strength of disorder.
Each line corresponds to an average over 25 realizations of noise (with $\alpha = 0$) on a lattice of
size $25\times90$.
The value of the condensate grows with increasing disorder strength, $w$.
Each line corresponds to a value of $\mu_0$ as indicated on the legend, but for the black dashed line which,
as explained in the text, marks the cut off used to define the critical temperature.}
\end{center}
\end{figure}

One of our main results  is the dependence of the condensate on the strength of the disorder, $w$, presented in Fig.
\ref{Cx-v-N}.
The results are qualitatively similar to the s-wave results \cite{Arean:2013mta}. 
As there, we observe that the average of the condensate grows as the strength of the disorder is increased. Moreover, for chemical potentials below the critical one, strong enough disorder drives the system into a phase where the average of the condensate is non-vanishing.

To make direct contact with the condensed matter literature we propose a disordered phase diagram in Fig. \ref{Tc}, where we track the value of the critical temperature of the normal to superconductor phase transition as a function of the strength
of the disorder. Let us explain, for the benefit of clarity how we have proceeded.
Since the value of the condensate increases with the the strength of the disorder, we have determined a value of the condensate above which we
consider the system in the superconducting phase (black dashed line in Fig. \ref{Cx-v-N}). We then read the average chemical potential\footnote{We refer to the average over realizations, not to be confused with the spatial average.} and
use the fact that the only relevant scale is $\mu/T$ to determine the critical temperature. Let us advance a potential
criticism to out method. Clearly, it would have been more relevant to compute the conductivities and determine the phase
diagram based on a conductivity criterion \cite{futureUS};
we expect, as in all previous cases, that there is a direct relation between the existence of a condensate and the transport
properties of the holographic solutions. One important aspect of Fig. \ref{Cx-v-N} is its robustness. Namely, the precise
form of the phase diagram varies quantitatively depending on where precisely we draw the cut off line defining the
``appearance'' of a nonzero condensate. However, qualitatively it is clear from the plot that the conclusions are stable
with respect to parallel shifts of the position of this cut off line.
\begin{figure}[htp]
\begin{center}
\includegraphics[width=3.5in]{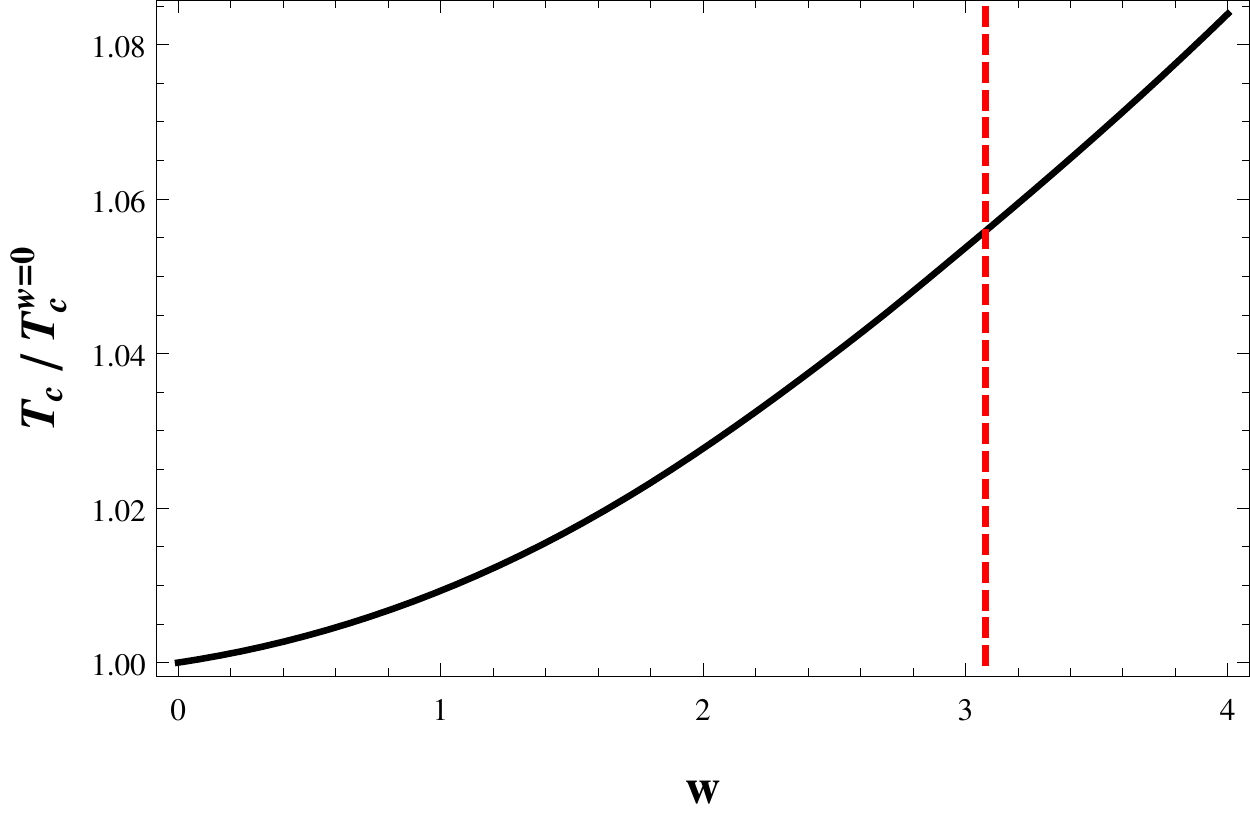}
\caption{\label{Tc} Enhancement of $T_c$ with the noise strength $w$ ($T_c^{w=0}$ stands for the critical
temperature in the absence of disorder). For values of $w$ to the right of the red dashed line the chemical potential becomes negative at its minimum.}
\end{center}
\end{figure}

Finally, let us try and explain the mechanism behind this enhancement of the critical temperature.
Looking at Eq. (\ref{meffwx}) it is not evident that the noise would enhance condensation by lowering the effective mass.
Actually the effect of the noise on the average (along $x$) of that effective mass is almost negligible.
However, the noise does have the effect of producing regions (in the $x$ direction) where the effective mass is below the
critical value for condensation. When these regions are large enough they trigger the condensation, resulting in solutions
where the condensate is nonzero along the whole sample (see Fig. \ref{ModelSimulationa02}); even in the regions where the chemical potential is below its critical value (for the homogeneous case) the condensate
is nonzero.

\section{Correlated noise}\label{Sec:Correlated}

In this section we shall analyze the case of correlated noise, namely that when in Eq. (\ref{noisefunc}) we consider 
$\alpha\neq 0$. 
Although, as we have seen in Eq.(\ref{xilargex}), this noise is correlated along the whole system, it is still worth looking at its effect on the condensate,
 and check if the main features of the response of the system are similar to those of the s-wave case studied in \cite{Arean:2013mta}.

Let us first specify the choice of parameters for our simulations. We will be using the function (\ref{noisefunc}) to implement a noisy chemical potential,
setting the power spectrum $\alpha=1.5$, the system length $L=2\pi$, and the UV cut off $k_* = 1/a_x$ saturating
the Nyquist limit. We again parametrize the strength of the noise in terms of $w={25\epsilon\over\mu_0}$, and
restrict $w$ to values for which the chemical potential $\mu(x)$ stays positive along the whole system.
We will run our simulations in lattices of size $N_z\times N_x=25\times75$, and use the numerical methods described  in section \ref{Subsec:disorder}.
\begin{figure}[hbt]
\begin{center}
\begin{subfigure}[b]{0.40\textwidth}
\includegraphics[width=\textwidth]{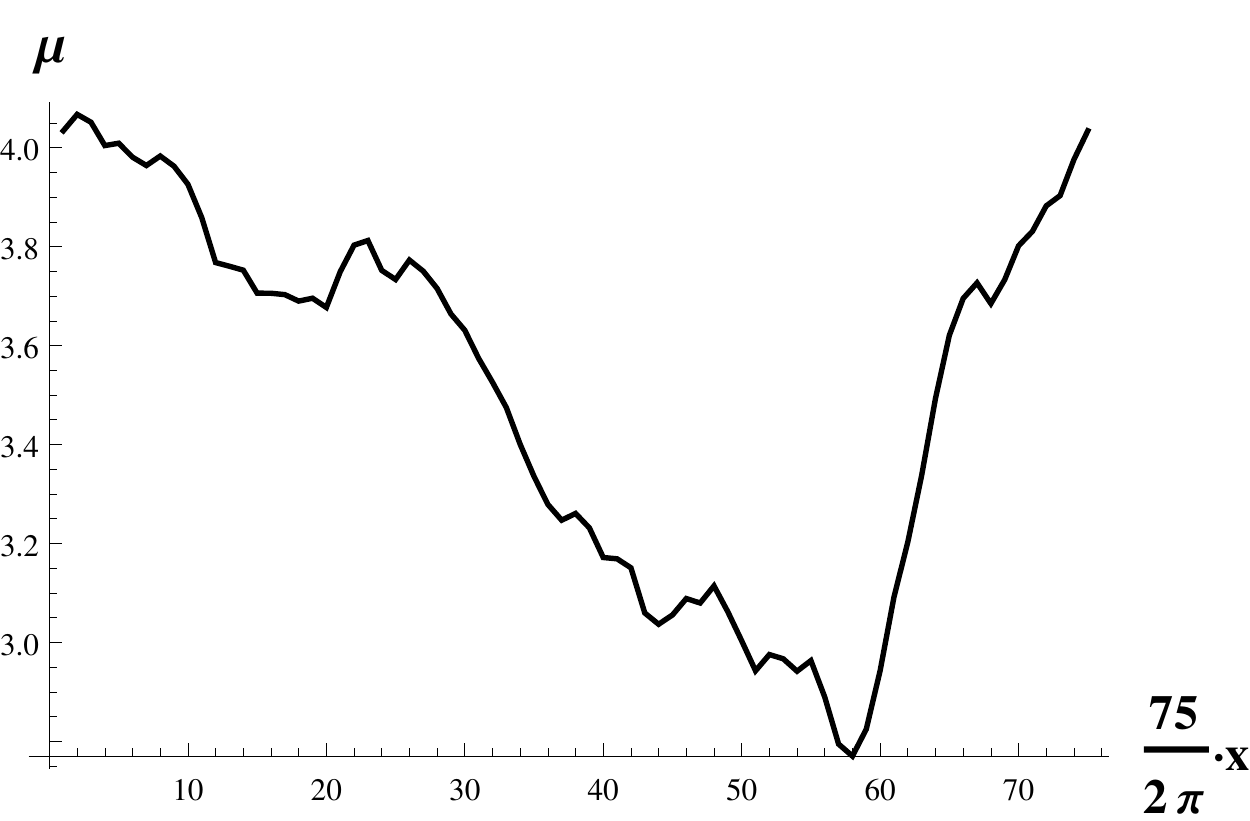}
\end{subfigure}
~
\begin{subfigure}[b]{0.40\textwidth}
\includegraphics[width=\textwidth]{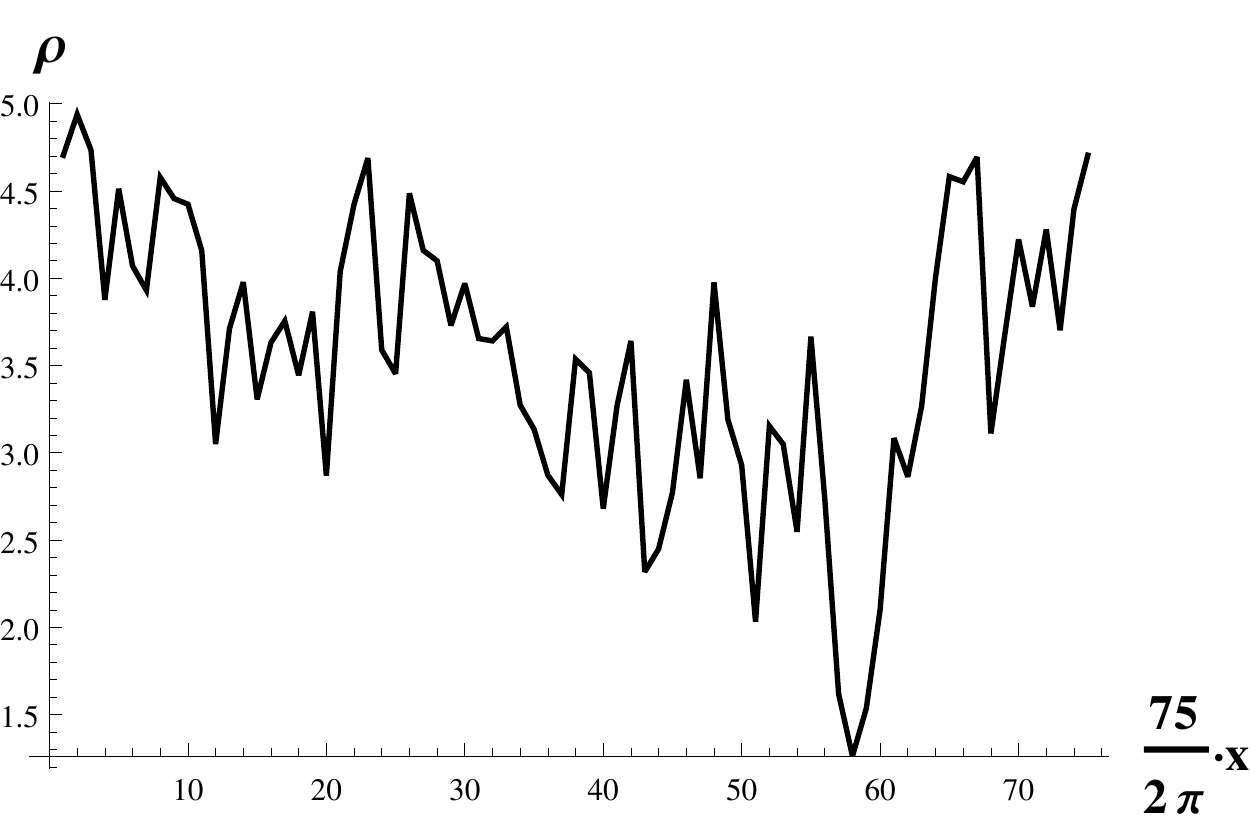}
\end{subfigure}\\[4mm]

\begin{subfigure}[b]{0.40\textwidth}
\includegraphics[width=\textwidth]{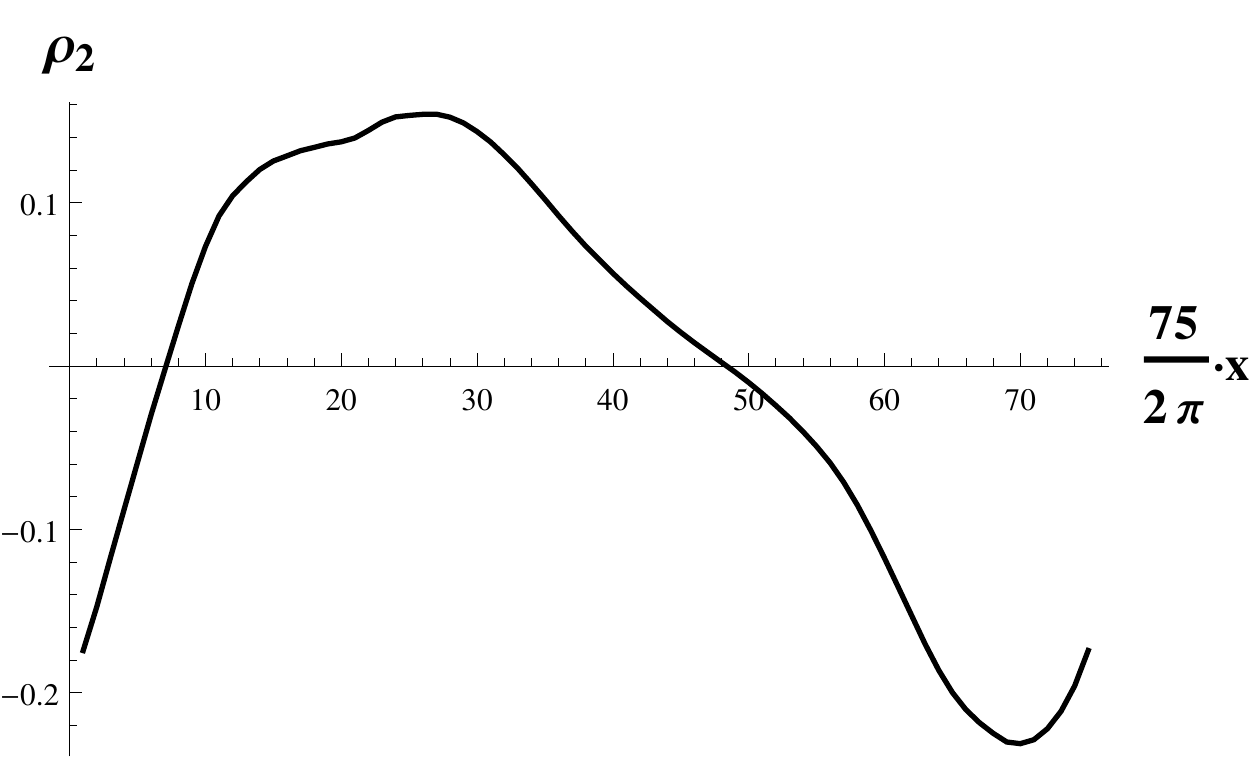}
\end{subfigure}
~
\begin{subfigure}[b]{0.40\textwidth}
\includegraphics[width=\textwidth]{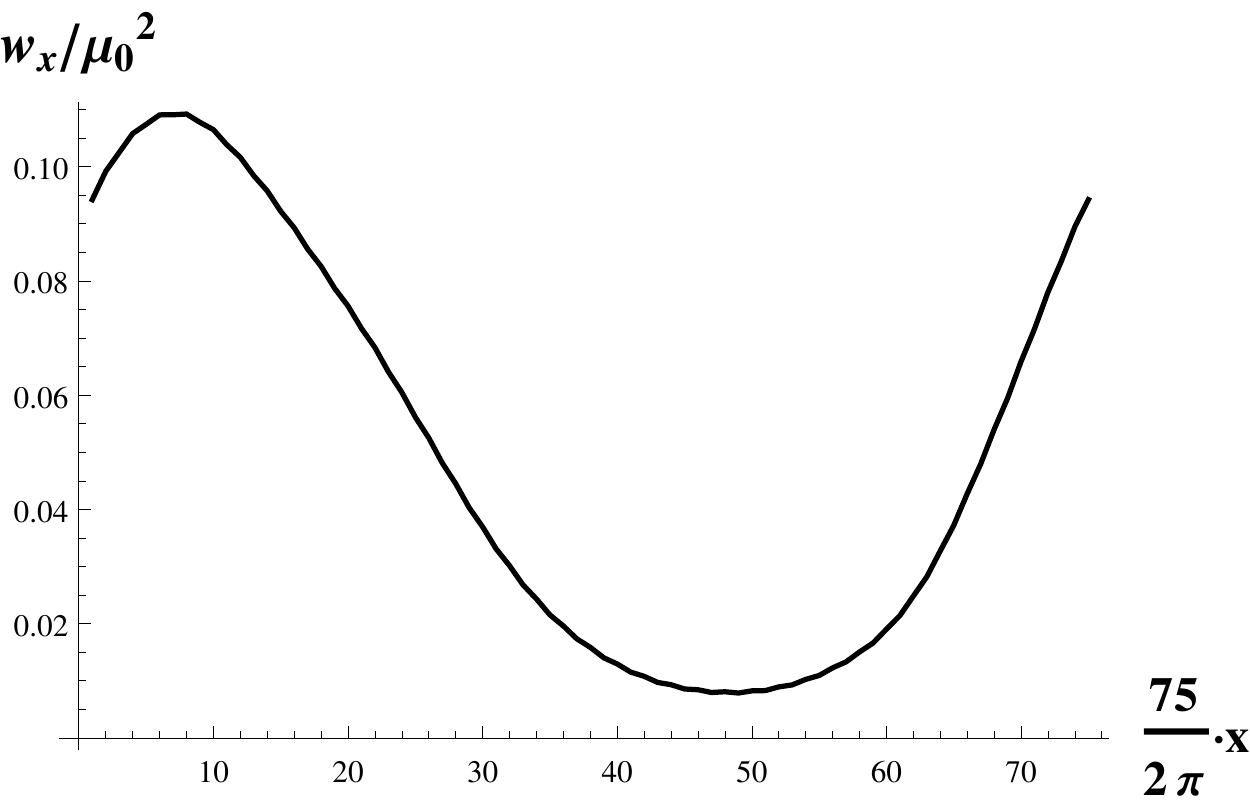}
\end{subfigure}
\caption{\label{ModelSimulationa15}
Example of a simulation corresponding to a noisy chemical potential with
$w=3.50$, $\alpha = 1.50$, and $\mu_0=3.50$ below the critical value of the homogeneous chemical potential
($\mu_c=3.66$). From the upper left panel and in clockwise sense: $\mu(x)$, $\rho(x)$,  $w_x(x)$,
and $\rho_2(x)$.}
\end{center}
\end{figure}

In figure \ref{ModelSimulationa15} we present a typical example of a simulation for a solution with correlated noise and chemical
potential $\mu_0=3.50$ below the critical value.
As one can see, the main features of this solution are similar to those of that in Fig. \ref{ModelSimulationa02} for the uncorrelated noise,
although some of the effects of the disorder are more pronounced in this case.
\begin{figure}[htb]
\begin{center}
\includegraphics[width=3.0in]{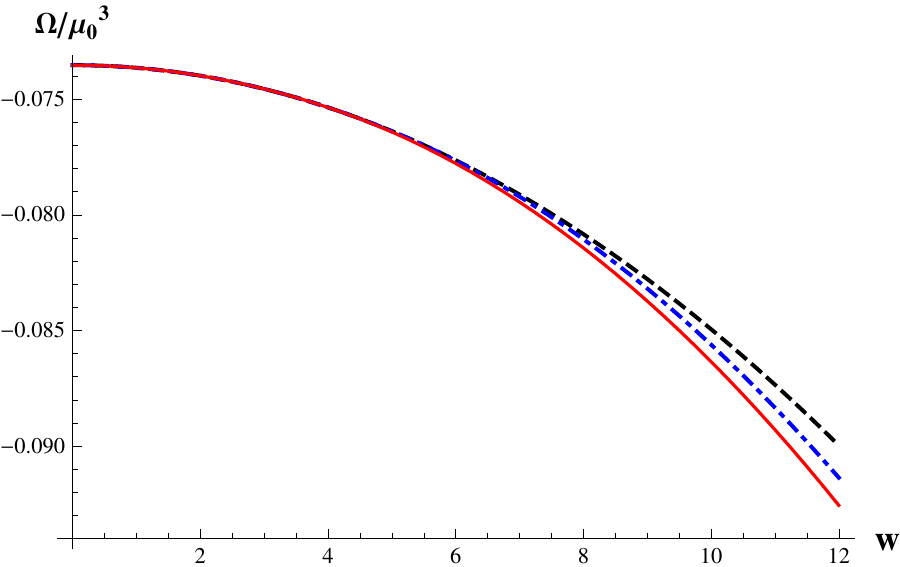}
\includegraphics[width=3.0in]{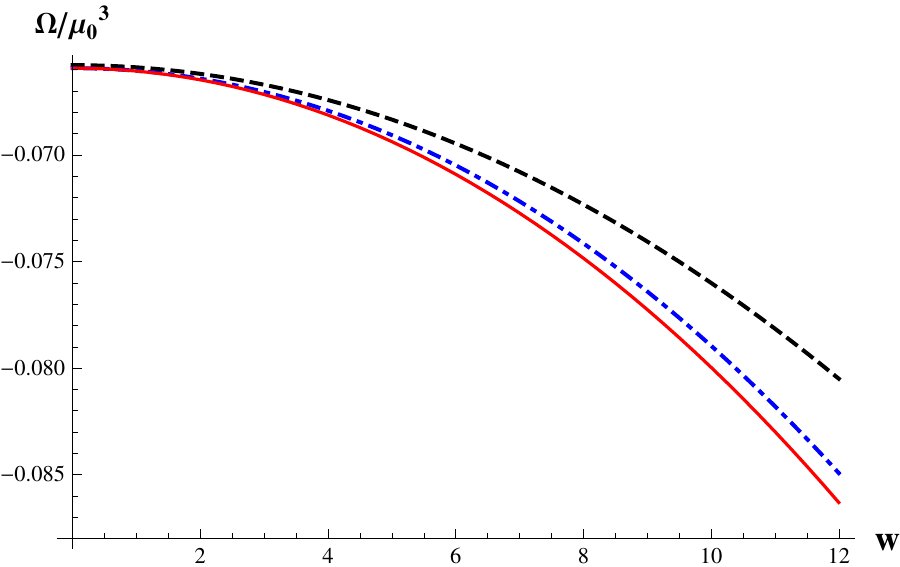}
\caption{\label{SvWcorr} Free energy of competing solutions as a function of the strength of the disorder,  $w$.
We consider values of  $\mu_0$ below and above the critical one.
The left panel corresponds to $\mu_0=3.4<\mu_c$, and the right one to $\mu_0=3.8>\mu_c$ (both with $\alpha=1.50$). The black dashed line corresponds to the normal phase solution, the blue dot-dashed line to the $Y$ solution, and the red solid line to the $X$ solution.
These plots result from averaging over 5 realizations on lattices of size $22\times40$.}
\end{center}
\end{figure}
For example, the difference in free energy among the different branches is
visible directly in the graphs in Fig.  \ref{SvWcorr}, where we plot the free energy of the X and Y condensed solutions plus that of the normal phase. Another interesting  observation following from that figure is that
for the panel corresponding to
$\mu_0=3.4<\mu_c$, all solutions coincide up to a noise strength $w\sim 6$. This reflects the fact that
for $w < 6 $ only the solution corresponding to the normal phase exists.

As in section \ref{Sec:Phase}, by repeating our simulations for different values of the average chemical potential we can study the dependence of the  condensate on the strength of the disorder.
The results are plotted in Fig. \ref{Cx-v-Ncorr}, and are qualitatively similar to those for the s-wave superconductor with correlated noise, presented in  \cite{Arean:2013mta}. There are, however, some differences.
\begin{figure}[hbt]
\begin{center}
\begin{minipage}{.6\linewidth}
\includegraphics[width=\linewidth]{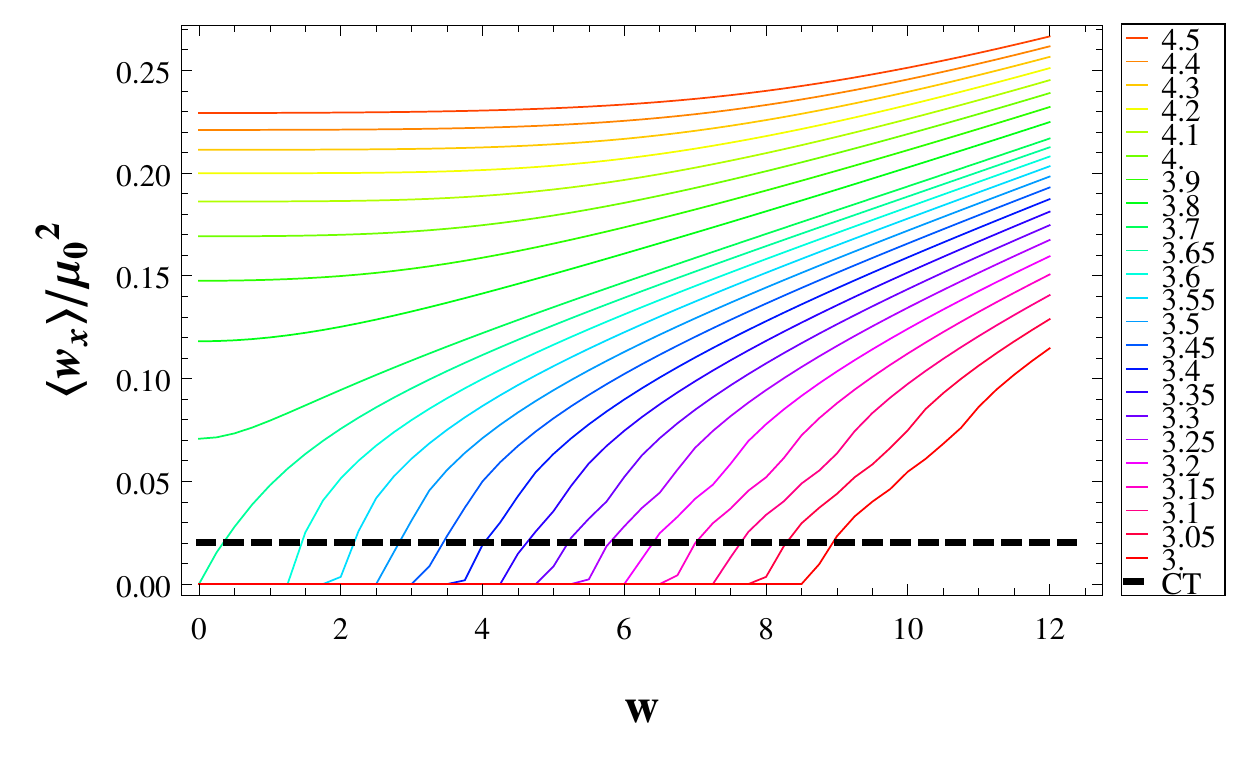}
\end{minipage}
\begin{minipage}{.39\linewidth}
\vspace{.4cm}
\includegraphics[width=\linewidth]{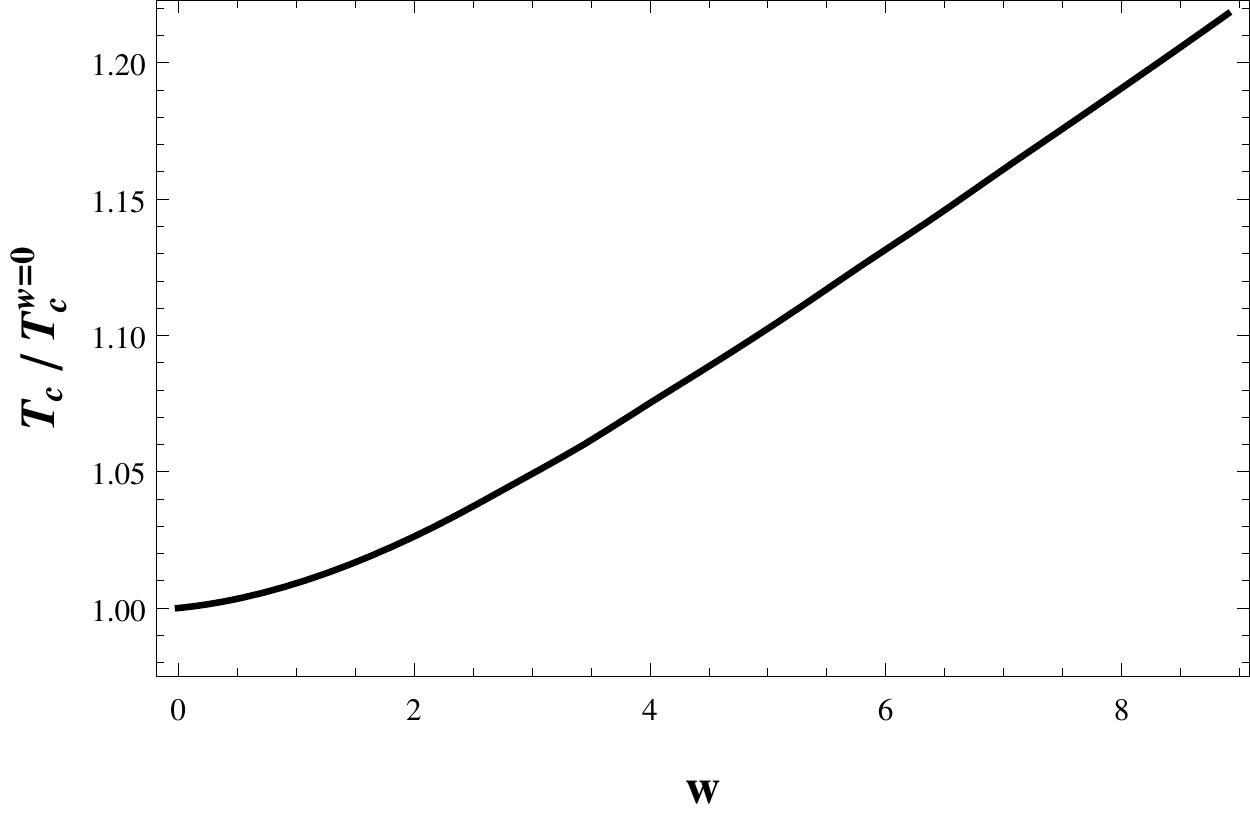}
\end{minipage}
\caption{\label{Cx-v-Ncorr} On the left we plot the spatial average of the condensate as a function of the
strength of disorder. Each line corresponds to an average over 10 realizations of noise (with $\alpha = 1.50$)
on a lattice of size $22\times40$.
Each line corresponds to a value of $\mu_0$ as indicated on the legend, but for the black dashed
line which marks the cut off used to define the critical temperature. On the right panel is shown the resulting
phase diagram.}
\end{center}
\end{figure}
First, the saturation for large values of $\mu_0$ is different as it seems that the condensate takes relatively larger values
of the average chemical potential with respect to the critical one to stabilize. This observation is somewhat marred by the
fact that, in general, going to very high values of the chemical potential leads to a region where back-reaction
has to be included and, therefore, renders the result in the probe limit unreliable. Second, there seems to be the case that larger
values of the chemical potential for the p-wave superconductor allow for an enhancement of the condensate with a higher
slope that larger values for the s-wave. There are, certainly, some similarities. For example, as in \cite{Arean:2013mta},
the curves are noisier at large disorder
strength and even noisier for lower values of the chemical potential. It could be suspected that such behavior is the result
of numerical limitations but we have run extensive simulations to that effect and verified that the effect is real and has to
do more with the nature of the system of equations in this regime of parameters.

Finally, as for the case of uncorrelated noise, a tentative phase diagram is presented on the right panel of figure \ref{Cx-v-Ncorr}. It is constructed from the data on the left panel of that figure, following the same procedure as that described in section \ref{Sec:Phase} for the uncorrelated noise. This phase diagram clearly shows that
the introduction of correlated noise results in an enhancement of the superconductivity very similar to that observed in \cite{Arean:2013mta} for the case of the s-wave superconductor.

\section{Spectral properties and disorder}\label{Sec:Spectrum}

In previous sections we have mostly focused on the average properties of the condensate and the charge density.
For example, in Figs. \ref{Cx-v-N} and \ref{Cx-v-Ncorr} we followed the average value of the condensate as a function of the strength of
the noise. 
Although this averaging is a good proxy at first, it is instrumental to the nature of disorder that we look  into
properties depending on the spatial coordinate $x$. In this section we will go beyond that first order averaging study.
Following \cite{Arean:2013mta} we continue the study of the spectral properties of some of the quantities characterizing
our system. Our goal is to gain a quantitative understanding.

As in \cite{Arean:2013mta}, we establish certain universality of the power spectra of the condensate  and charge density
as functions of the power spectrum of the signal defining the noise. Namely, for a given random signal with power spectrum
of the form $k^{-2\alpha}$ we study the power spectrum of the condensate  $k^{-2\Delta(\alpha)}$, of the charge density
$k^{-2\Gamma(\alpha)}$, and of the $\rho_2$ charge density $k^{-2\Gamma_2(\alpha)}$; and report some interesting
universal behavior. We interpret this behavior as a particular form of renormalization of small wave-lengths.
We argue that this kind of smoothing/roughening points to a renormalization of sorts, where higher harmonics in ${\cal O}$
are suppressed or enhanced with respect to their spectral weight in $\mu$.

\begin{figure}[b]
\begin{center}
\includegraphics[width=.99\textwidth]{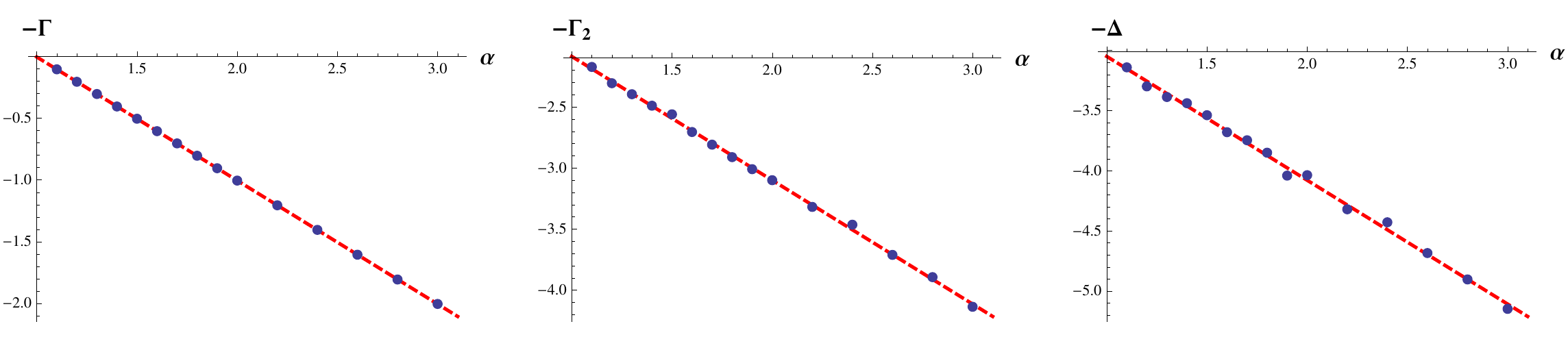}
\caption{\label{RGToy} Renormalization of the disorder: Charge density $\rho$: $\Gamma= -1.00+ 1.00\,\alpha $ (left panel),  charge density $\rho_2$: $\Gamma_2= 1.07+ 1.01\,\alpha $ (middle panel), and condensate:
$\Delta=2.02+1.03\,\alpha$ (right panel).
This plot was made considering $L=2 \pi$, $\mu_0=4$, $w=1$, and averaging over 5 realizations on a lattice of size
$22\times75$.
}
\end{center}
\end{figure}

Let us now carefully describe our setup.
To characterize this renormalization quantitatively we take a boundary chemical potential of the form presented in
equation (\ref{noisefunc}), but now considering different values for $\alpha$ (the choice of $\alpha$ determines
the degree of differentiability -smoothness- of the initial profile).
To make the concept of renormalization more precise we will study the power spectrum of the modes characterizing the response
of our system (the condensate, charge density $\rho$, and charge density $\rho_2$), as a function of the input power spectrum
determining our noise.
This input power spectrum of $\mu(x)$ is essentially proportional to $k^{-2\alpha}$. Remarkably, the power spectrum of the
condensate ${\cal O}(x)$ is numerically well approximated by $k^{-2\Delta}$ (see the rightmost panel of Fig.
\ref{RGToy}). We find $\Delta\simeq 2.02+1.03\alpha$, which is clearly larger than $\alpha$, meaning that
the weight of the high-$k$ harmonics is smaller in ${\cal O}$ than in $\mu$.
The power spectrum of the charge density $\rho$ is very well approximated by $k^{-2\Gamma(\alpha)}$ (Fig.
\ref{RGToy} left panel) with
$\Gamma \simeq -1.00+1.00 \alpha$, which implies that for the charge density the weight of the high-$k$ harmonics is
larger than in the spectrum of $\mu$. As for the charge density $\rho_2$, again the spectrum approximates very well to
a power law $k^{-2\Gamma_2(\alpha)}$ (Fig. \ref{RGToy} middle panel), with $\Gamma_2\simeq 2.02+1.03 \alpha$. As for
the condensate the weight of the high-$k$ harmonics is smaller than in $\mu$.

Let us stress that the spectra of all the response functions are given by power laws, and moreover, the
exponent of these response power laws is always of the form $\sim \alpha + {\rm integer}$.
This universality of RG is one of the main observations of our work and its origin seems to be in the strongly coupled
nature of the problem. The weak field theory intuition would dictate that $\Delta$ should be well approximated by
the conformal dimension associated with the order parameter and here we verify that it is not.

It is also interesting to point out that this behavior does not depend on any of the parameters of our theory, i.e.
$L$, $T/\mu$ or $w$. This means that we can redo Fig. \ref{RGToy} for the charge density in the normal phase.
This particular case is interesting, since the theory
becomes linear and we can therefore separate variables.
Being that the case, we can recompute the power spectrum solving the equations of motion using a
simple $Mathematica$'s NDSolve command. In this case we get $\Gamma= -1.00+ 1.00\,\alpha $,
which agrees with the result presented in Fig. \ref{RGToy}.
It is worth mentioning that the same scalings were found in \cite{Arean:2013mta} for the $s$-wave holographic superconductor,
and similar ones have been observed in a related holographic model
by other authors in \cite{Hartnoll:2014cua}.

\section{Conclusions}\label{Sec:Conclusions}

In this paper we have studied the influence of disorder in the holographic p-wave superconductor. We have found that
moderate disorder enhances the value of the order parameter and accordingly of the critical temperature.
We have also discussed various branches of solutions that appear in this particular setup since different solutions are
characterized by the dominant direction of the condensate.

We have established that the dominant solution, according to its free energy, is the one with the condensate along the
$x$-direction. We have also established that the condensate, $\langle {\cal O}_x\rangle/\mu^2$, is enhanced with the
disorder. Moreover, we have demonstrated the self-averaging property of $\langle {\cal O}_x\rangle$ under Gaussian
and uncorrelated randomness.  We identify this enhancement with the ulterior enhancement of superconductivity.
The phase diagram is similar to the s-wave superconductor reported in \cite{Arean:2013mta} and we presented its
quantitative form in section \ref{Sec:Phase}. The key property is that the curve delimiting the normal and
superconducting phase shows an enhancement of the superconductivity with mild disorder. We have also studied some
universal properties of the power spectrum of the corresponding condensate and charge density.
We have found that the response is largely governed by a simple linear relation depending on the power spectrum of
the random chemical potential. These results expand those presented first in  \cite{Arean:2013mta} to the case of a
disordered holographic p-wave superconductor. Similar behavior was also reported in \cite{Hartnoll:2014cua} in the
perturbative regime for a neutral scalar, and by some of the authors in gravity duals of brane intersections
\cite{futureUS}.
One of our main results, the enhancement of superfluidity with mild disorder,
is  aligned with experimental and numerical claims of p-wave superfluidity enhancement by disorder
\cite{PhysRevB.79.214529,0295-5075-86-2-26004}.

We would like to finish by highlighting a few problems that are particularly interesting to us and some of which we hope
to pursue in future works. Having constructed the disordered solutions in \cite{Arean:2013mta} and the present manuscript,
it is natural to study transport properties and, in particular, the conductivities.  It would also be interesting to
understand the effects of disorder in more general types of holographic p-wave superconductors. Recall that this type of
superconductors present a particularly interesting challenge to Anderson's theorem given its directional order parameter.
Some interesting models include \cite{Gubser:2008wv} and its extension to $p_x+ip_y$ along the lines of
\cite{Zayas:2011dw}. A recent study of conductivity in p-wave superconductors was presented in  \cite{Herzog:2014tpa},
where some phenomenological  similarities with high temperature cuprate superconductors were found even in a translational
invariant holographic model. It would be interesting to study the persistence or modification of such properties under the
effect of introducing disorder as a way of breaking translational invariance in these and similar systems.

Finally, as in \cite{Arean:2013mta}, we have established the existence of fairly universal response of the condensate and the charge density to the
power spectrum of the random disorder. We view this as evidence of some universality in cases of strongly coupled
systems under a sort of disorder renormalization. On the other hand, for very small values of $k_0$, we found evidence of a
new scaling for the expectation values of one point functions. We expect to study this potential universality in more detail
in the future \cite{futureUS}.

\section*{Acknowledgments}
We would like to thank M. Ara\'ujo, J. Sonner and T. Takayanagi for useful discussions.
D.A, L.A.P.Z. and I.S.L. thank the Abdus Salam ICTP, Italy for hospitality at various stages of this project.
I.S.L. thanks  Max-Planck-Institut f\"ur Physik for hospitality.
D.A. thanks the FRont Of pro-Galician Scientists for unconditional support.
We also thank the Bivio;  for being the epic place we needed to finish this project.
Some simulations
were performed in the University of Michigan Flux high-
performance computing cluster. This work is partially
supported by Department of Energy under grant DE-
FG02-95ER40899. The work of D.~Are\'an is supported by GIF, grant 1156

\bibliographystyle{JHEP}
\bibliography{CMbib}

\end{document}